\documentclass[prb,twocolumn,aps,amssymb,footinbib,showpacs]{revtex4}
\usepackage{amssymb}
\usepackage{graphicx}
\usepackage{amsmath}
\usepackage{times}
\usepackage{color}
\usepackage{subfigure}
\usepackage{setspace}
\usepackage{bm}

\begin{document}

\newcommand {\beq} {\begin{equation}}
\newcommand {\eeq} {\end{equation}}
\newcommand {\bqa} {\begin{eqnarray}}
\newcommand {\eqa} {\end{eqnarray}}
\newcommand {\da} {\ensuremath{d^\dagger}}
\newcommand {\ha} {\ensuremath{h^\dagger}}
\newcommand {\adag} {\ensuremath{a^\dagger}}
\newcommand {\no} {\nonumber}
\newcommand {\ep} {\ensuremath{\epsilon}}
\newcommand {\ca} {\ensuremath{c^\dagger}}
\newcommand {\ga} {\ensuremath{\gamma^\dagger}}
\newcommand {\gm} {\ensuremath{\gamma}}
\newcommand {\up} {\ensuremath{\uparrow}}
\newcommand {\dn} {\ensuremath{\downarrow}}
\newcommand {\ms} {\medskip}
\newcommand {\bs} {\bigskip}
\newcommand{\kk} {\ensuremath{{\bf k}}}
\newcommand{\rr} {\ensuremath{{\bf r}}}
\newcommand{\kp} {\ensuremath{{\bf k'}}}
\newcommand {\qq} {\ensuremath{{\bf q}}}
\newcommand{\nbr} {\ensuremath{\langle ij \rangle}}
\newcommand{\ncap} {\ensuremath{\hat{n}}}

\begin{abstract}
  We theoretically study the dynamic screening properties of bilayer
  graphene within the random phase approximation assuming quadratic
  band dispersion and zero gap for the single-particle spectrum. We
  calculate the frequency dependent dielectric function of the system
  and obtain the low energy plasmon dispersion and broadening of the
  plasmon modes from the dielectric function. We also calculate the
  optical spectral weight (i.e. the dynamical structure factor) for
  the system. We find that the leading order long wavelength limit of
  the plasmon dispersion matches with the classical result for 2D
  electron gas. However, contrary to electron gas systems, the
  non-local plasmon dispersion corrections decrease the plasmon
  frequency. The non-local corrections are also different from the
  single layer graphene. Finally, we also compare our results with the
  double layer graphene system (i.e. a system of two independent
  graphene monolayers).

\end{abstract}

\title{Dynamic Screening and Low Energy Collective Modes in Bilayer Graphene }
\date{\today}
\author{  Rajdeep Sensarma, E. H. Hwang, and S. Das Sarma}  
 \affiliation{ Condensed Matter Theory Center, Department of Physics, University of Maryland, College Park, USA 20742}
\maketitle

\section{Introduction}

Bilayer graphene (BLG) has recently attracted experimental and
theoretical interest, both for fundamental physics as well as for
possible technological
applications~\cite{blg1,blg2,hwang:rossi,blgtransport:expt,optics:expt,slgreview}.
The system consists of two layers, each of which has carbon atoms arranged in
a honeycomb lattice with two (A and B) sublattices. Each layer can be
viewed as a single layer graphene (SLG)~\cite{slgreview} with their
associated physics of chiral Dirac Fermions, coupled by strong
interlayer tunneling between A and B lattice sites in the two
layers. It is useful to view BLG as an effective layer where the
interlayer tunneling changes the linear dispersion of the
quasiparticles in SLG to an effective quadratic dispersion at small
wavevectors \cite{blg2}. 
Thus, a
simple model for BLG single-particle spectrum is a pair of chiral
parabolic electron and hole bands touching each other at the Dirac (or
the charge neutrality) point. Each band has a 4-fold degeneracy
arising from spin and valley degrees of freedom. The pristine or
undoped bilayer graphene has attracted a lot of interest since the
presence of a single Fermi point and quadratic dispersion can lead to
a host of exotic phenomena~\cite{kunyang1,levitov1}. The low energy
properties of the doped bilayer graphene also show interesting
features like enhanced backscattering due to the chirality of the
bands present in the system~\cite{hwang:blgstatic}. The
single-particle spectrum is sufficiently different in SLG (linear
dispersions) and BLG (quadratic dispersions) that it is theoretically
interesting to explore and contrast various dielectric properties in
the two systems. In this paper we consider the dynamic screening and
low energy collective modes of BLG.

Coulomb interaction and its dynamic screening due to many-body effects
is an important property of any electronic material. While static
screening determines transport properties, the dynamic (frequency
dependent) screening determines the elementary quasiparticle spectra
and the collective modes and is crucial to understanding the optical
properties of the system. While much work has been done on screening
properties and plasmon modes of SLG both
theoretically~\cite{hwang:slgdynamic} and
experimentally~\cite{slgexpt}, there are very few existing analytic
works on doped bilayer systems. There are numerical calculations of
the bilayer dynamic screening within a four-band
model~\cite{macdonald}, and the parabolic
approximation~\cite{chakravorty1,chakravorty2}, but it is not useful
in a general context. Ref.~\onlinecite{chakravorty2} obtains results which
are qualitatively similar to the results in this paper. Recently, the
static screening properties of BLG systems have been studied
analytically in Ref.~\onlinecite{hwang:blgstatic}. In this paper, we
will analytically study the dynamic screening of Coulomb interactions
in doped BLG and obtain its collective modes.

The parabolic dispersion makes the BLG low energy physics quite
different from SLG systems. On the other hand, although the system has
the same low energy dispersion as a regular two dimensional electron
gas (2DEG) system, the chirality of graphene leads to features which
are distinct from a standard 2DEG~\cite{2DEG} which has been
extensively studied in the context of semiconductor
heterostructures~\cite{semiconductor}. It is therefore useful to
compare the physics of BLG with SLG and 2DEG. In this respect, it is
important to note that although bilayer graphene consists of two
layers of graphene sheets, it should really be viewed as a 2D system
with parabolic dispersion and chiral bands for the purpose of low
energy physics. Thus, it is also important to compare this system with
other bilayer systems such as double quantum
wells~\cite{Madhukar,Hwang:tunnel}, which consists of two layers of
2DEG coupled by Coulomb interaction and shows distinct features as a
result of its quasi-2D nature. Another closely related system is the
double layer graphene, which is obtained by putting a layer of oxide
between two single layers of graphene\cite{dlg:expt}. In this case,
the layer separation is much larger ($\sim 10 nm$) than in BLG ($\sim
0.3 nm$) and the system is better viewed as two different layers of
graphene coupled by Coulomb interactions without any interlayer
tunneling. The dynamic screening properties of BLG is quite different
from these systems. It is then useful to compare the results obtained
for BLG with corresponding results in these two bilayer systems
(i.e. two layers of 2DEG and two layers of SLG).


\begin{figure*}[t!!]
\includegraphics[scale=0.2]{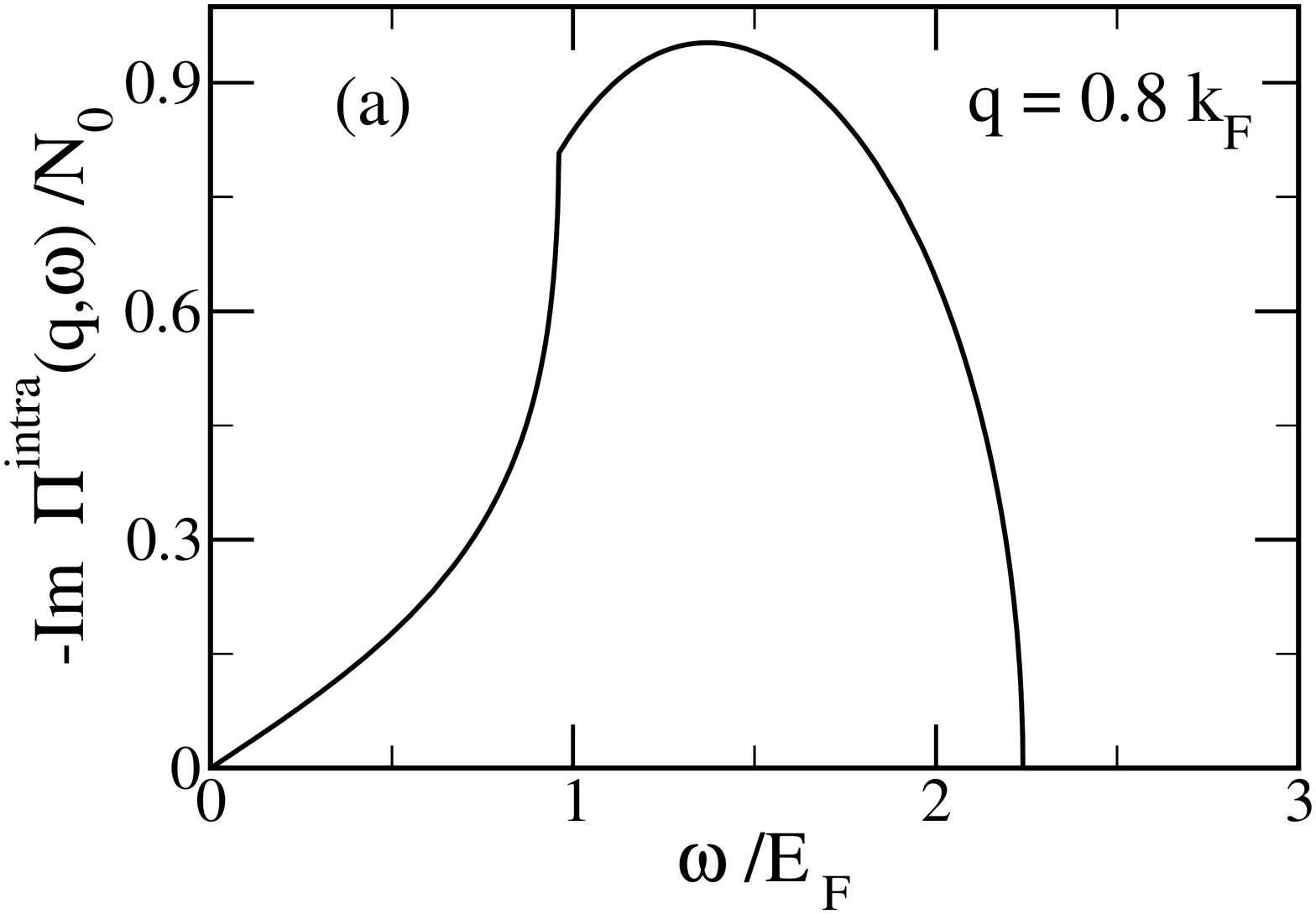}
\includegraphics[scale=0.2]{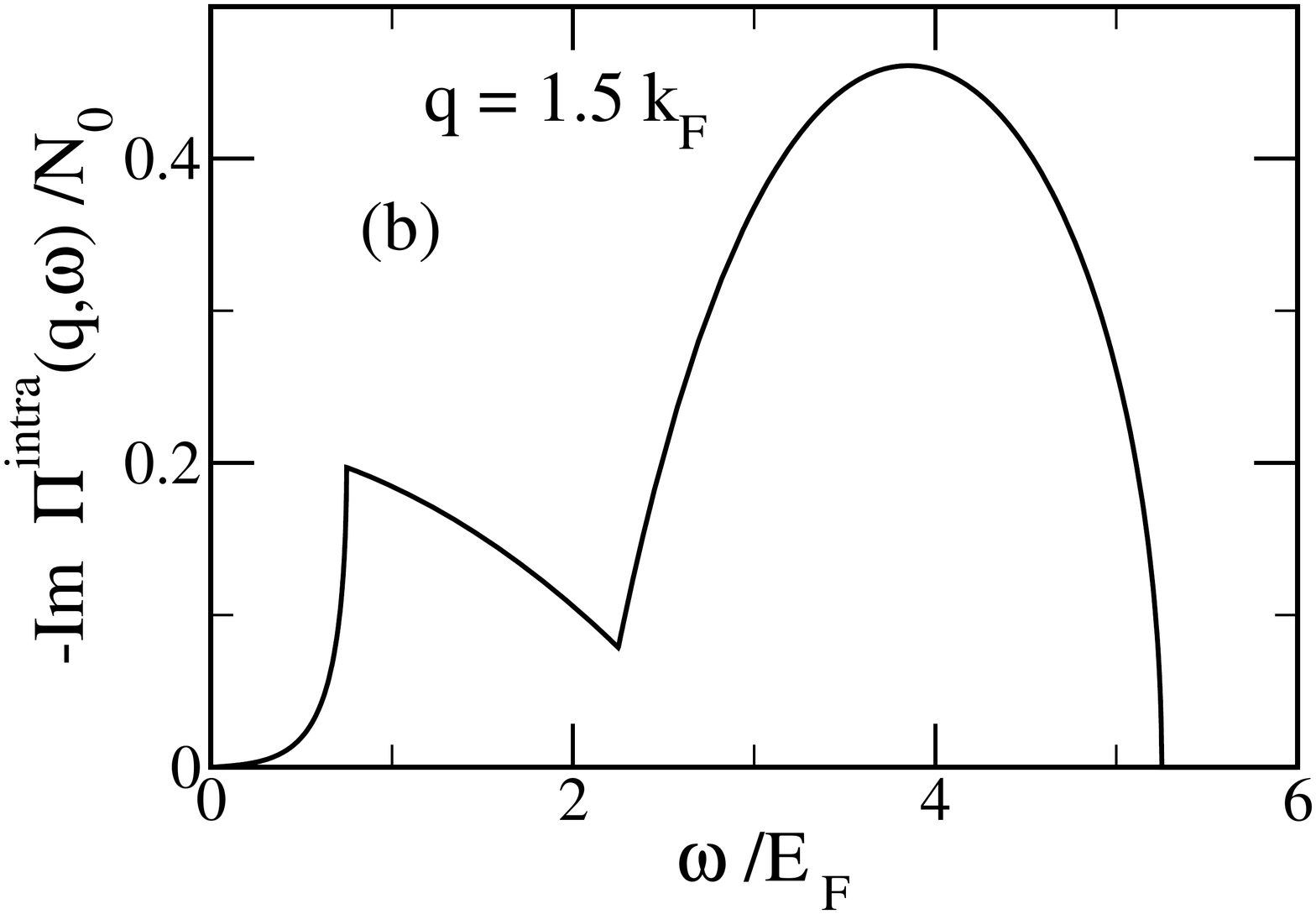}
\includegraphics[scale=0.2]{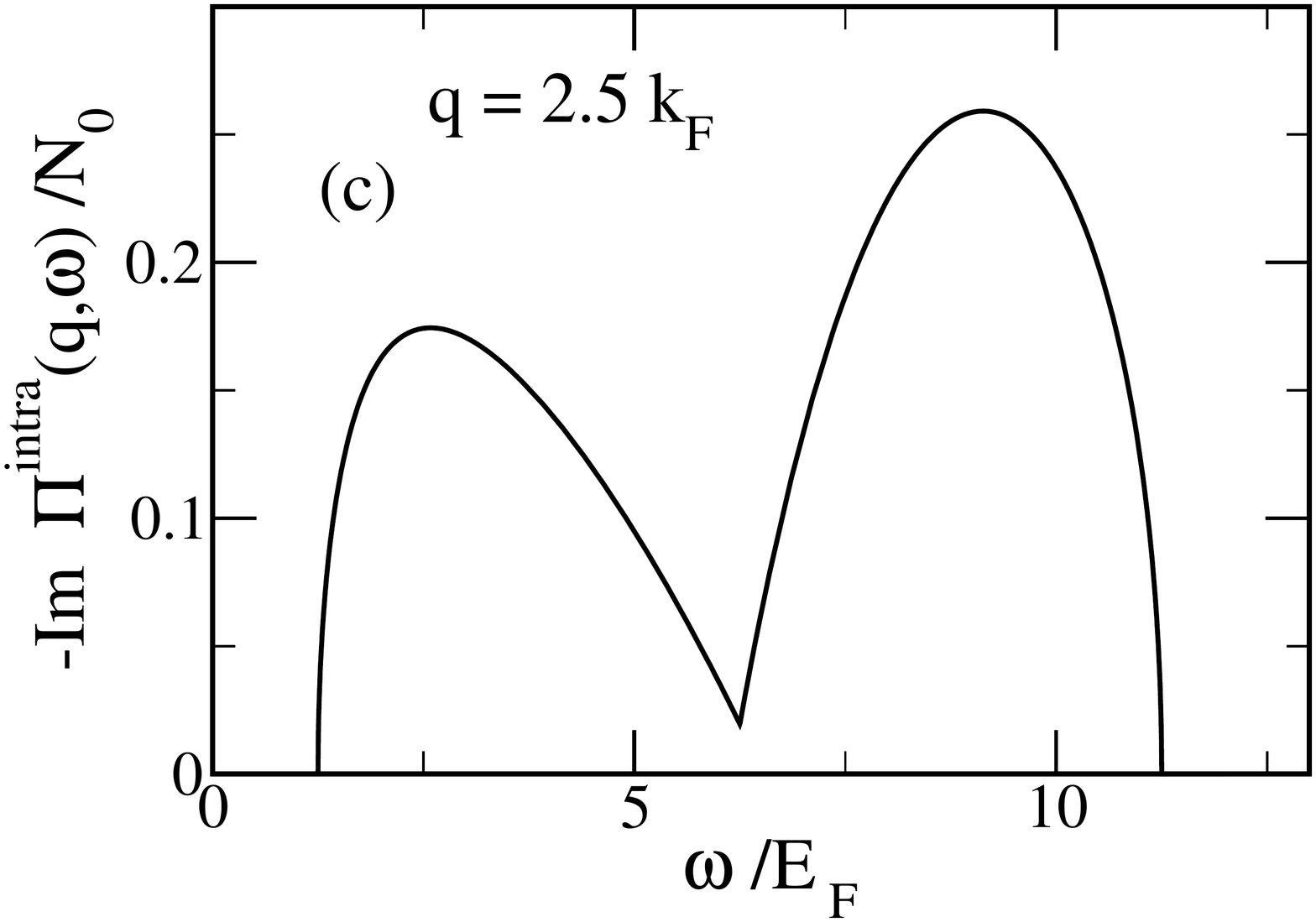}
\includegraphics[scale=0.2]{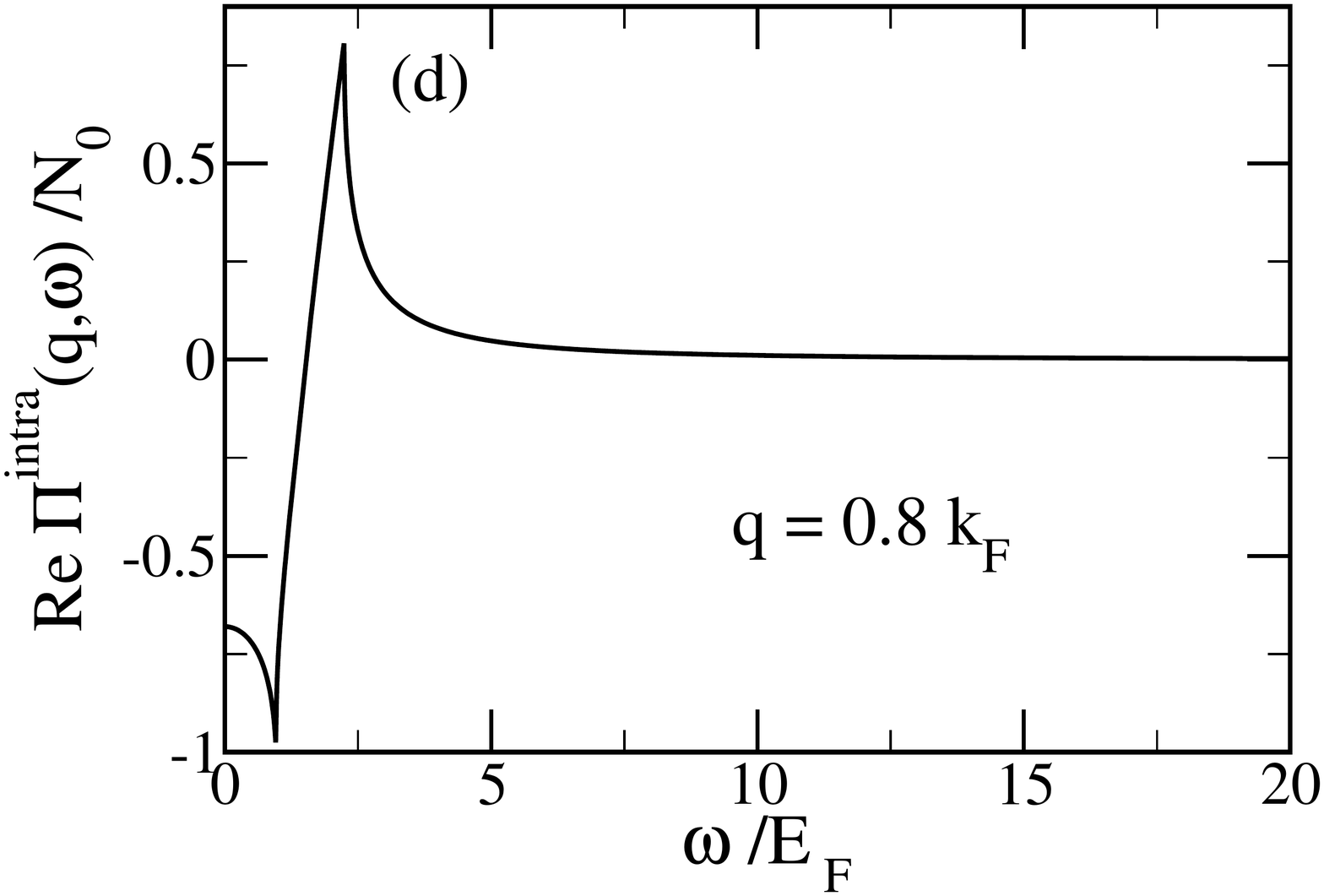}
\includegraphics[scale=0.2]{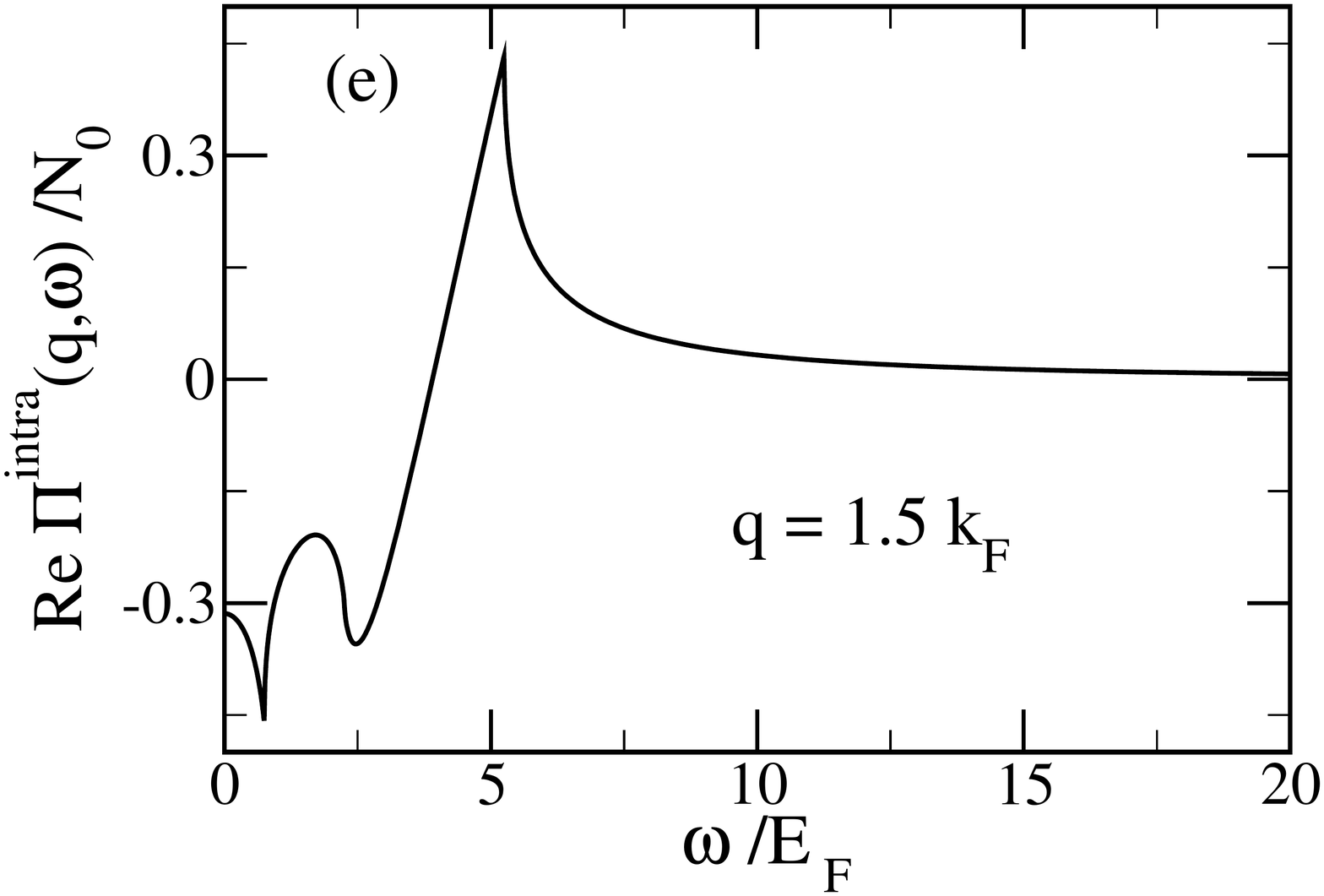}
\includegraphics[scale=0.2]{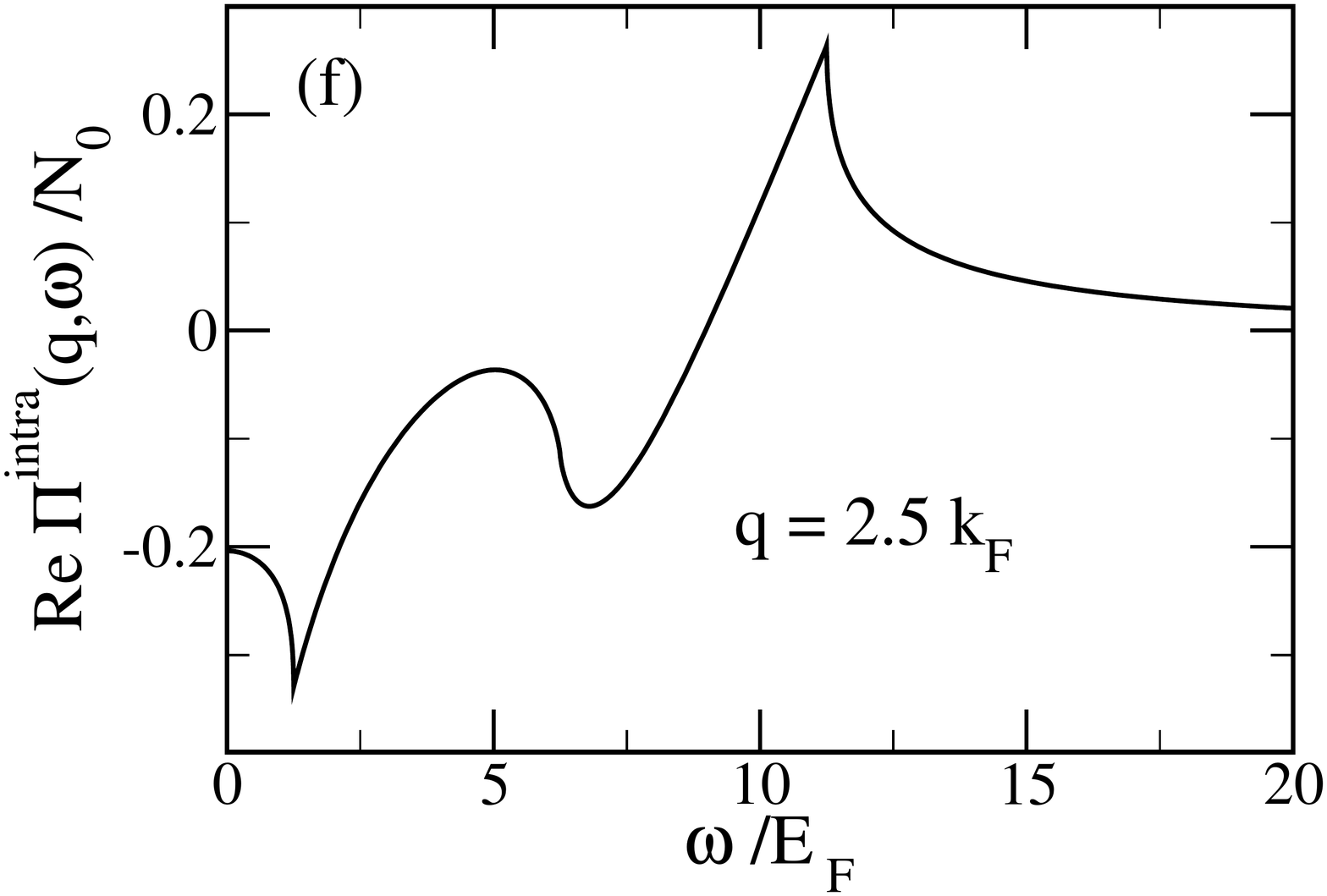}
\caption{(a)-(c): Imaginary part of Intraband susceptibility as a function of
  frequency for (a) $q=0.8 k_F$ (b) $q=1.5 k_F$ and (c) $q=2.5 k_F$. (d)-(f): Real part of Intraband susceptibility as a function of
  frequency for (d) $q=0.8 k_F$ (e) $q=1.5 k_F$ and (f) $q=2.5 k_F$.}
\label{polintra}
\end{figure*}

In this paper, we analytically study the dynamic screening properties
of the Coulomb interaction in BLG systems within the random phase
approximation (RPA). We calculate the dielectric function of bilayer
graphene, $\ep(\qq,\omega)$ at arbitrary wavevectors $\qq$ and
frequency $\omega$. The imaginary part of $1/\ep(\qq,\omega)$ is
related to the optical spectral weight which can be directly measured
in optical conductivity \cite{li2008} or light-scattering
\cite{lightscatter}, or electron-scattering measurements
\cite{slgexpt,ieel} while the zeroes of the real part give the
dispersion of the plasmon modes. We calculate the plasmon modes and
their damping due to the presence of the second band in the system. We
make extensive comparisons of our results with similar results for SLG
systems~\cite{hwang:slgdynamic} and 2DEG~\cite{2DEG}, especially
regarding the dependence of collective mode dispersions on the carrier
densities in these systems. We look at the density dependence of the
long wavelength collective mode dispersion and show how the system
crosses over from a BLG like behaviour to a SLG like behaviour as
density of the carriers is increased. We also compare our results with
those for quasi-2D systems like double quantum wells~\cite{Madhukar}
and double layer graphene~\cite{hwang:dlg} and show that in spite of
having two layers, bilayer graphene shows unique features in its
dynamic screening properties.

We note that we are working here with the parabolic approximation to
the dispersion of the BLG bands. In real bilayer systems, the
dispersion changes from parabolic to a linear form (i.e. the full
dispersion is hyperbolic) at high energies. In addition the 2 band
approximation breaks down at higher energies, resulting in a 4-band
BLG model. However, it is clear that for low density materials, the
low energy properties of the system are qualitatively well described
by keeping only the parabolic part of the dispersion. For bilayer
graphene systems, our parabolic approximation for collective modes
should be valid upto densities $\sim 5\times 10^{12} cm^{-2}$.

Our BLG model, as mentioned above, consists of a 2-band (single
valence and conduction band) gapless parabolic chiral 2D
single-particle energy dispersion. We compare our analytic dynamic
screening and collective mode BLG results
with SLG (chiral gapless 2-band linear dispersion), 2DEG (non-chiral
1-band gapped parabolic dispersion), double-layer graphene (two
parallel SLG) and 2D bilayer ( two parallel 2DEG) systems. Our
theoretical results, therefore, give a fairly complete picture of
dielectric screening and plasmon modes in 2D systems,
covering both single and bilayer systems, both linear and quadratic
band dispersions, system with and without gaps, and both chiral and
non-chiral situations. Our calculated BLG plasmon dispersion can be
directly compared with experimental results when they become
available.

The low energy effective Hamiltonian for BLG is given by 
\beq
H=-\frac{1}{2m}\left( \begin{array}{cc}%
0 & (k_x-ik_y)^2 \\
(k_x+ik_y)^2 & 0 
\end{array} \right),
\label{eqn:ham}
\eeq
where $m=\gamma/(2v_F^2)$ , $\gamma$ is the interlayer tunneling
amplitude inherent in the BLG system and $v_F$ is the graphene Fermi
velocity. We set $\hbar=1$ throughout this paper. Typically, in
the BLG systems $m\simeq 0.033 m_0$, with $m_0$ being the free
electron mass. This Hamiltonian differs
from the SLG case in having a quadratic as opposed to a linear
dispersion. As a result, the system has a constant density of states
rather than the linear density of states in SLG. As we will see, this leads to 
substantial difference in the BLG dielectric response.

The Hamiltonian can be diagonalized to obtain two bands with
dispersions $\ep^s_\kk=s\kk^2/(2m)$ and corresponding wavefunctions
$\Psi^s_\kk =\frac{e^{i\kk\cdot \rr}}{\sqrt{2}}\left(e^{-2i\chi_\kk},s
\right)$, where $s= \pm 1$ denotes respectively the conduction and the
valence band.  Here $\chi_\kk=\tan^{-1} (k_y/k_x)$. In this paper we
use the parabolic Hamiltonian to obtain the dynamic screening within
random phase approximation (RPA). This involves a theoretical calculation of the dynamical dielectric function $\epsilon(\qq,\omega)$.
 
\begin{figure*}[t!]
\includegraphics[scale=0.2]{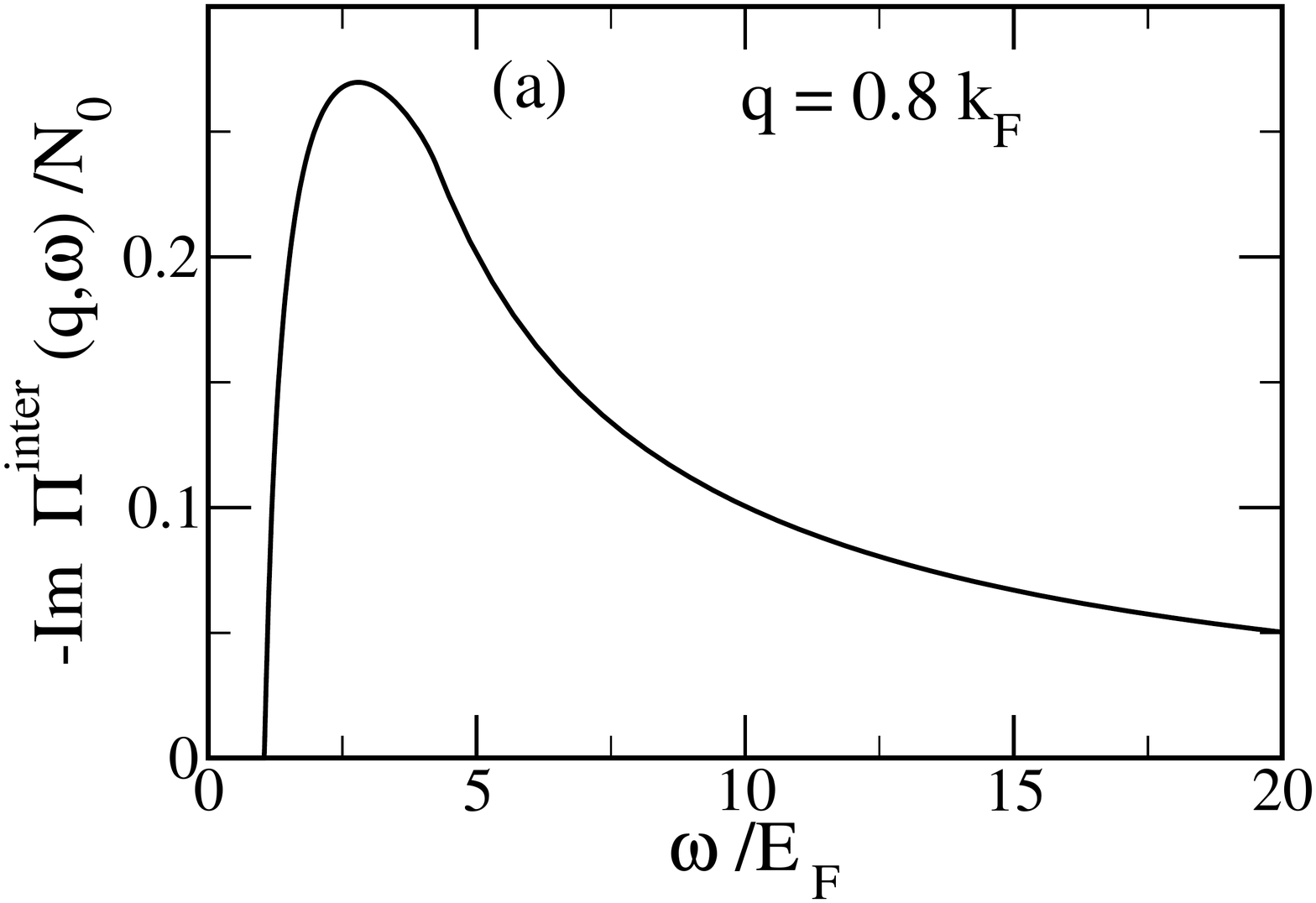}
\includegraphics[scale=0.2]{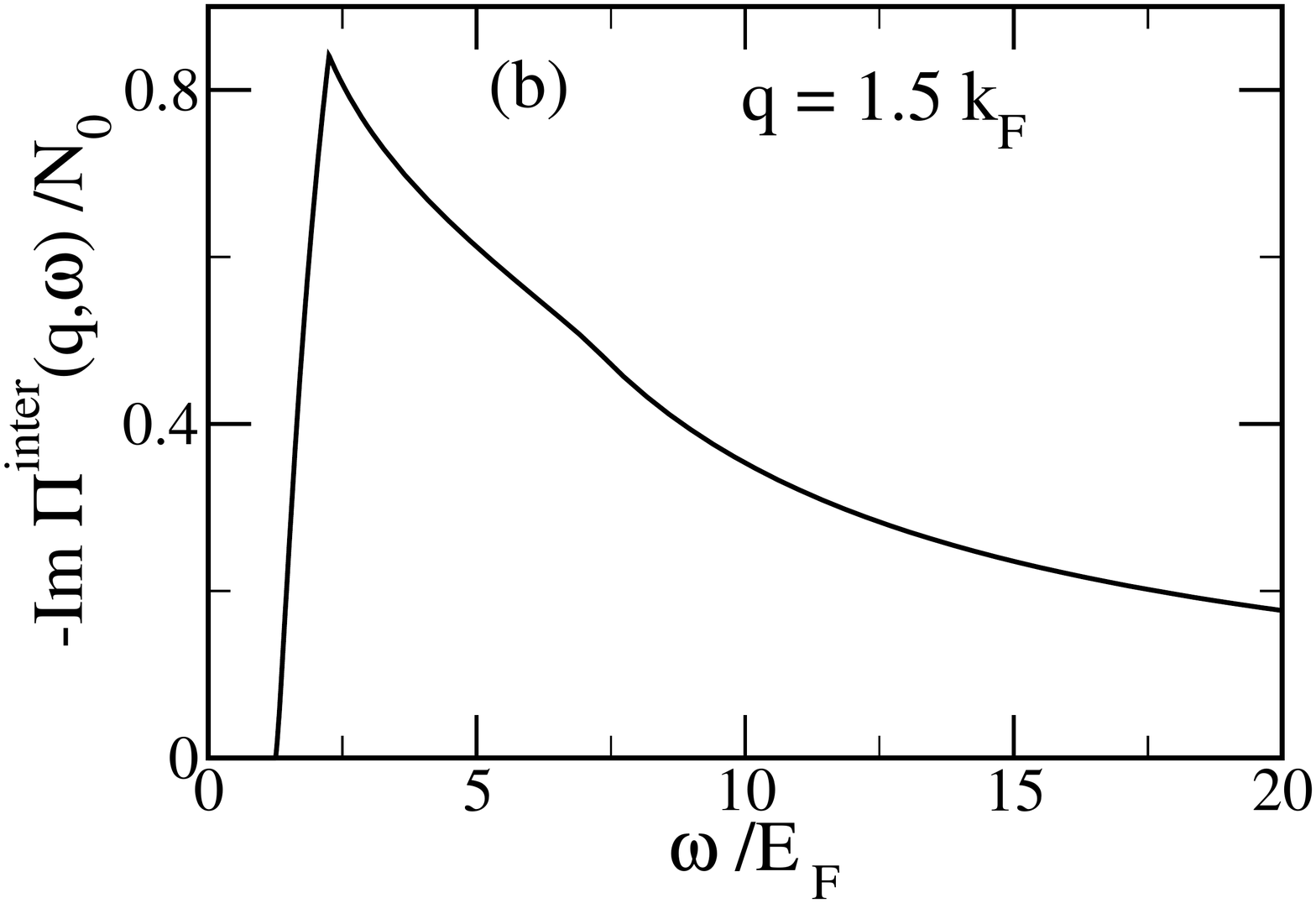}
\includegraphics[scale=0.2]{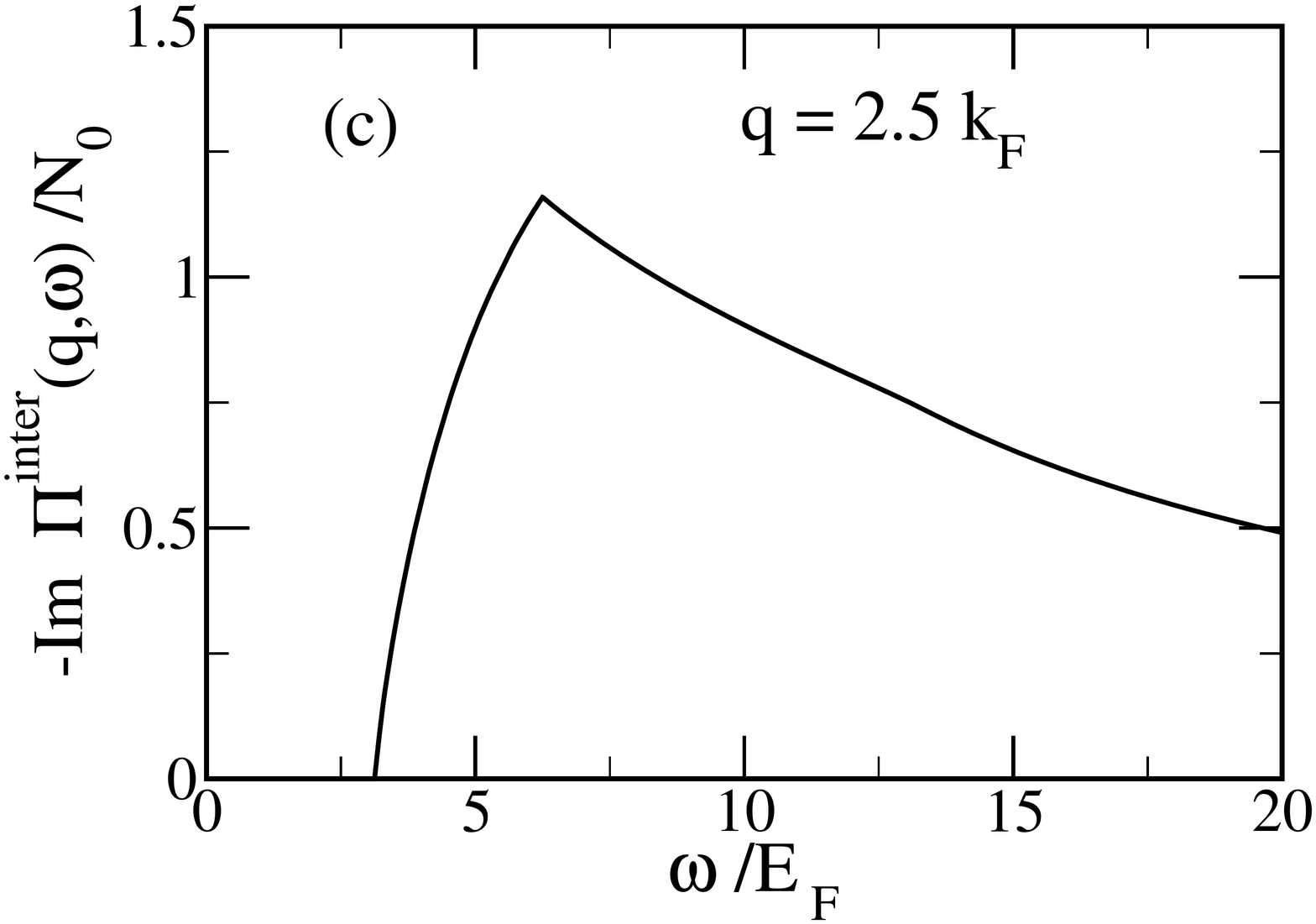}
\includegraphics[scale=0.2]{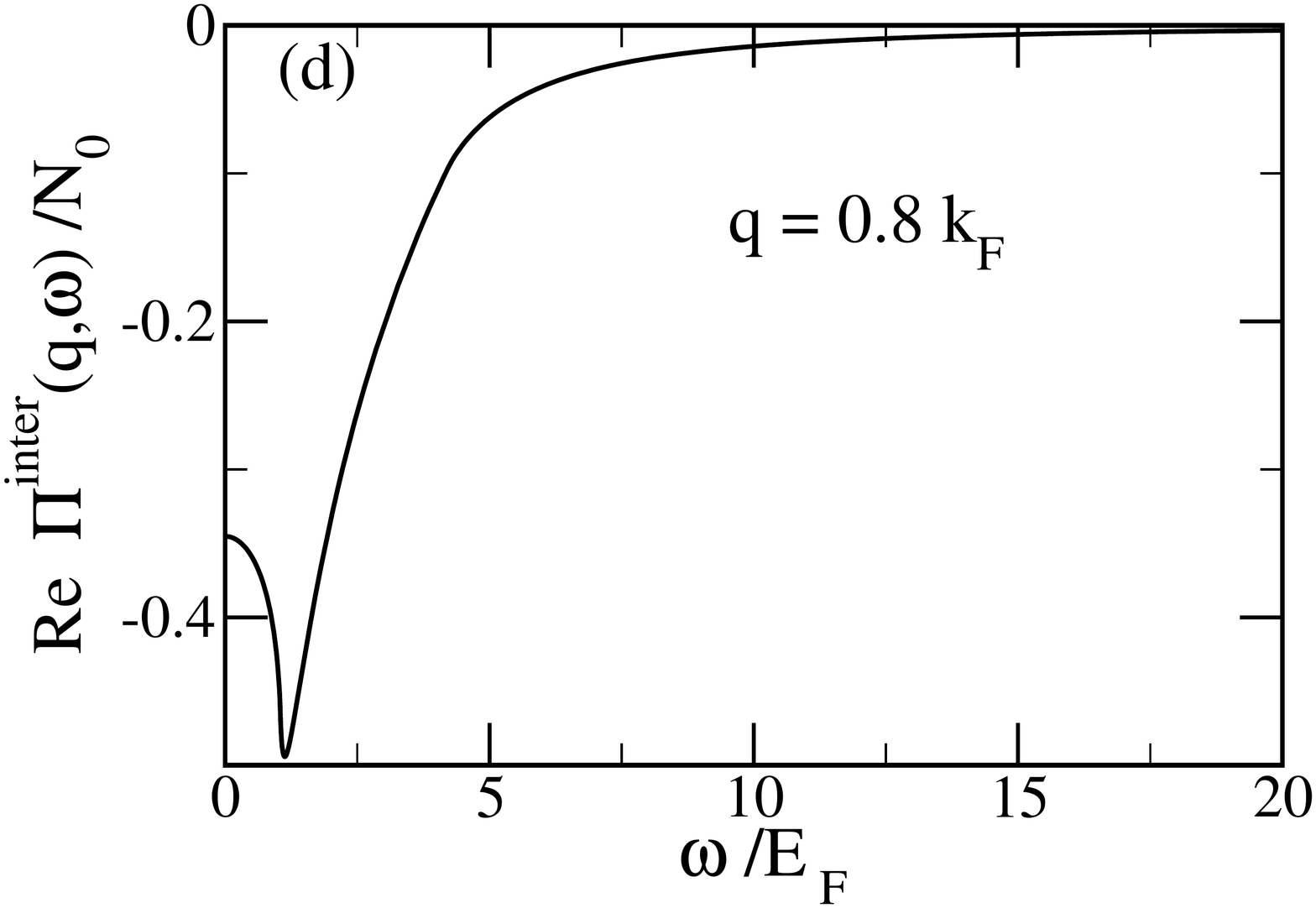}
\includegraphics[scale=0.2]{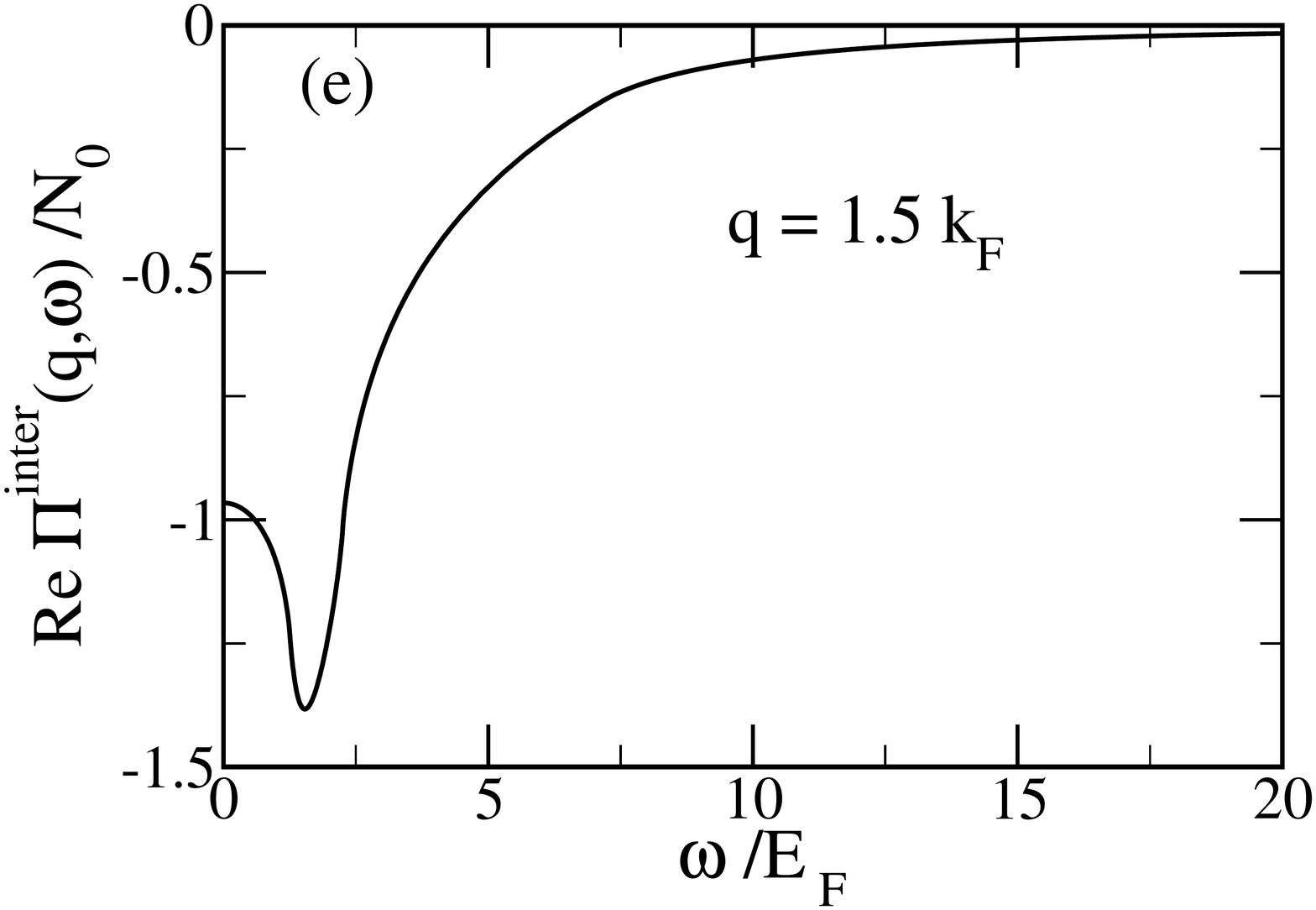}
\includegraphics[scale=0.2]{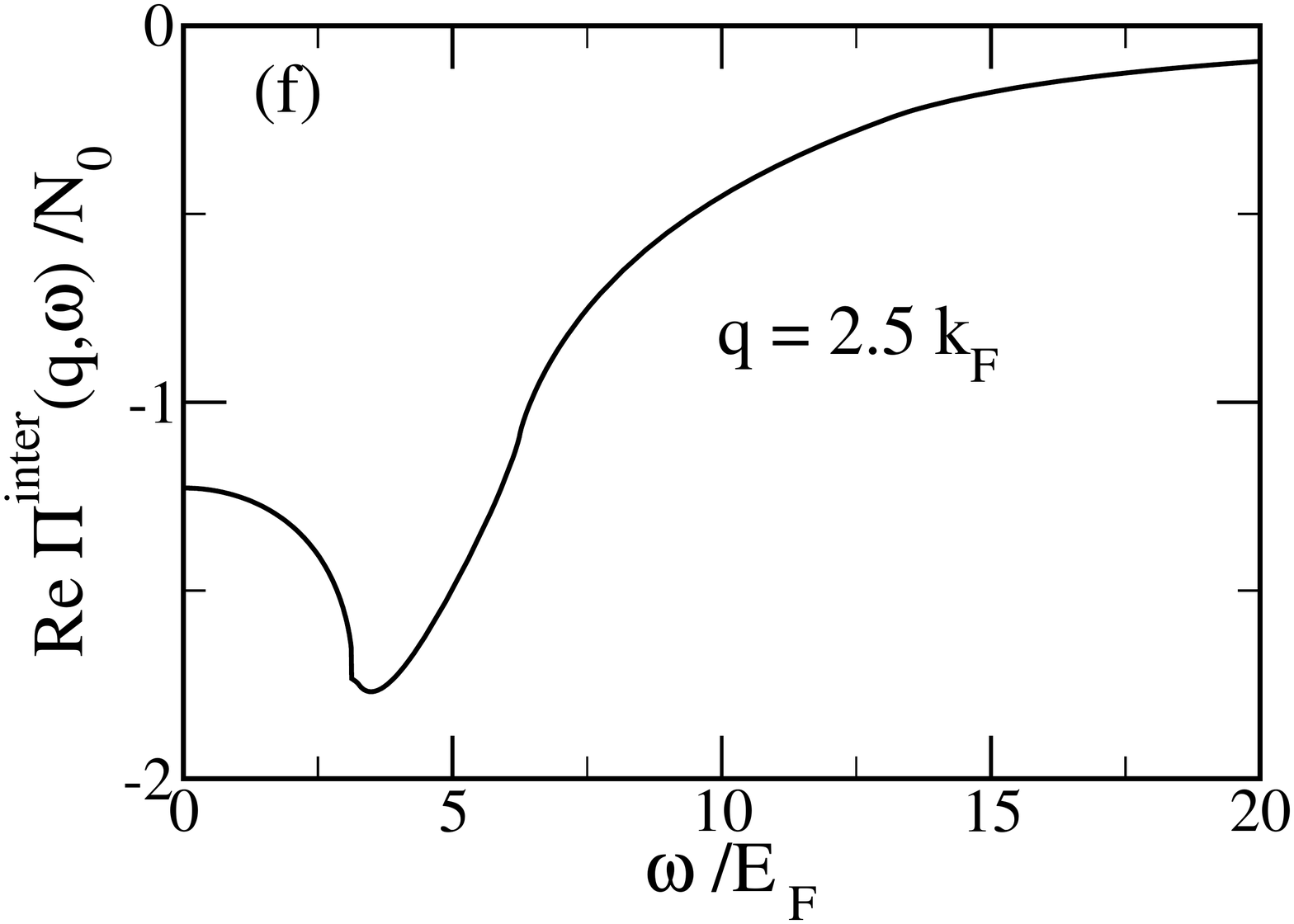}
\caption{(a)-(c): Imaginary part of Interband susceptibility as a
  function of frequency for (a) $q=0.8 k_F$ (b) $q=1.5 k_F$ and (c)
  $q=2.5 k_F$. (d)-(f): Real part of Interband susceptibility as a
  function of frequency for (d) $q=0.8 k_F$ (e) $q=1.5 k_F$ and (f)
  $q=2.5 k_F$.}
\label{polinter}
\end{figure*}
\section{Bare Bubble Polarizability}
The frequency dependent dielectric function for this system (within RPA)
is given by
\beq
\epsilon(\qq,\omega)= 1-\frac{2\pi e^2}{\kappa q}\Pi(\qq,\omega), 
\label{eqn:dielectric}
\eeq
where $\kappa$ is the background dielectric constant. Here $\Pi$ is
the free particle (bare bubble) polarizability, given by
\beq
\Pi(\qq,\omega)=g \sum_{ss'\kk}\frac{f^s_\kk-f^{s'}_{\kk+\qq}}{\omega+\ep^s_\kk-\ep^{s'}_{\kk+\qq}}F_{ss'}(\kk,\kk+\qq)
\label{eqn:barebubble},
\eeq
where $f^s_\kk$ is the Fermi function, $g$ is the degeneracy factor ($g=4$) and the wavefunction overlap
$F_{ss'}(\kk,\kk+\qq)=\frac{1}{2}[1+ss' \cos (2\theta)]$
with $\theta$ being the angle between $\kk$ and $\kk+\qq$. Using a
circular co-ordinate system $(k,\phi)$, one can write $\cos
\theta= (k+q\cos \phi)/|\kk+\qq|$. The chirality effect is 
incorporated in the matrix element $F$.

The polarizability can be separated into the {\it intraband} ($s=s'$
terms) and the {\it interband} ($s'=-s$ terms) contributions. The
scale for the polarizability is set by the free particle density of
states $N_0=gm/(2\pi)$. Using the Fermi momentum $k_F^{-1}$ and the
Fermi energy $E_F$ as units of length and energy respectively, we work
with the dimensionless variables $x=k/k_F$, $y=q/k_F$ and
$z=\omega/E_F$. Then, we can write
$\Pi(q,\omega)=N_0[\Pi^{inter}(y,z)+\Pi^{intra}(y,z)]$, where
$\Pi^{intra}$and $\Pi^{inter}$ are dimensionless quantities. Note that
the interband contribution is important in graphene only because of
its gapless nature; in semiconductor based 2DEG, the interband
contribution is implicitly included in the background dielectric
constant and is not considered in the electronic polarizability.

We consider a gated or doped BLG with $E_F\neq 0$, where the Fermi
level lies in the conduction band. At $T=0$, this implies that
$f^+_\kk=\Theta(k_F-k)$ and $f^-_\kk=1$. Then one can write
$\Pi^{intra}(y,z)=\Pi^1(y,z)+\Pi^1(-y,-z)$, where
\bqa  
\no \displaystyle \Pi^{1}(y,z)& = &\frac{1}{\pi}\int_0^1xdx\displaystyle{\int_{-\pi}^\pi}d\phi\frac{1}{z-y^2-2xy\cos \phi}\\
 & & ~~~~~~~~~~\left[1-\frac{y^2\sin^2 \phi}{x^2+y^2+2xy\cos \phi}\right],
\label{eqn:pi1}
\eqa
and  $\Pi^{inter}(y,z)=\Pi^2(y,z)+\Pi^2(-y,-z)$, where
\bqa  
\no \displaystyle \Pi^{2}(y,z)& = &\frac{-1}{\pi}\int_1^\infty xdx\displaystyle{\int_{-\pi}^\pi}d\phi\frac{\sin^2 \phi}{x^2+y^2+2xy\cos \phi}\\
 & & ~~~~~~~~~~~~\frac{y^2}{z+2x^2+y^2+2xy\cos \phi}.
\label{eqn:pi2}
\eqa
The azimuthal integrals in eqns. (\ref{eqn:pi1}) and (\ref{eqn:pi2}) can be done
analytically to obtain 
\bqa
\Pi^1(y,z)& = &\frac{-1}{2}\displaystyle \int_0^1\frac{dx}{x(x^2+z)}\left[ x^2+z-|x^2-y^2|\right.\\
\no &- & \left. Sgn[z-y^2+2xy]\frac{(2x^2+z-y^2)^2}{\sqrt{(z-y^2)^2-4x^2y^2}}\right],
\label{azint:pi1}
\eqa
\bqa
\no\Pi^2(y,z)&=&\frac{1}{2}\displaystyle \int_1^\infty \frac{dx}{x(x^2+z)}\left[x^2+z+|x^2-y^2|\right.\\
 & -&  Sgn[(x-y)^2+x^2+z]\\
\no & &\left.~~~ \sqrt{(2x^2+z+y^2)^2-4x^2y^2}\right],
\label{azint:pi2}
\eqa
where $Sgn(x)= \pm 1$, depending on $x \gtrless 0$. The real and imaginary parts of the polarizability are even
and odd under $z \rightarrow -z$; so we will only consider the case of
$z >0 $.  The real and imaginary parts of the intraband polarizability
are 
\bqa
\no \text{Re}~ \Pi^{intra}(y,z)& =&f_1(y,z)+f_1(y,-z)\\
\no &- &\Theta[(z-y^2)^2-4y^2]f_2(y,z)\\
 &- &\Theta[(z+y^2)^2-4y^2]f_2(y,-z),
\label{repi:intra}
\eqa
\bqa
\no \text{Im}~ \Pi^{intra}(y,z)& =&-\Theta[4y^2-(z-y^2)^2]f_3(y,z)\\
& + &\Theta[4y^2-(z+y^2)^2]f_3(y,-z).
\label{impi:intra}
\eqa
For notational brevity, it is useful to define the quantities
$a=(z-y^2)/2y$ and $b=(z+y^2)/2y$. Then we have
\bqa
\no f_1(y,z)&=& \frac{ya}{2z}\log\left\vert \frac{4y^2a^2}{ 1-z}\right\vert-\frac{1}{2}\\
& + &\frac{ya}{z}\Theta[1-y]\log\left\vert\frac{1-z}{2ya}\right\vert
\label{eqn:f1},
\eqa
\bqa
\no f_2(y,z)&=& Sgn[a]\left[\frac{\sqrt{a^2-1}}{y}+\frac{yb}{2z}\log\left\vert \frac{\sqrt{a^2-1}-b}{\sqrt{a^2-1}+b}\right\vert\right.\\
& - &\left.\frac{ya}{2z}\log\left\vert\frac{\sqrt{a^2-1}-a}{\sqrt{a^2-1}+a}\right\vert\right],
\label{eqn:f2}
\eqa
\bqa
\no f_3(y,z)&=&\frac{\sqrt{1-a^2}}{y}+\frac{y|a|}{z}\tan^{-1}\left[\frac{\sqrt{1-a^2}}{|a|}\right]\\
 & - &\frac{y|b|}{z}\tan^{-1}\left[\frac{\sqrt{1-a^2}}{|b|}\right].
\label{eqn:f3}
\eqa
%
For the interband polarizability, it
is useful to define: $c=y\sqrt{y^2+2z}$ and
$d=y\sqrt{2z-y^2}$. Then we can write
\bqa
 \text{Re}~\Pi^{inter}(y,z) =& f_4(y,z)+f_4(y,-z)+f_5(y,z)\\
\no  +&\Theta[(z-2)^2+d^2]Sgn(2-z-|d|)f_5(y,-z) 
\label{repi:inter}
\eqa
\bqa 
\no \text{Im}~\Pi^{inter}(y,z)& =  &\Theta[2z-y^2]\Theta[z-2+d]\\
\no & & \left[\Theta[z-2-d]\frac{\pi}{2}\left(\frac{|z-y^2|}{z}-1\right)\right.\\
&  &\left.+\Theta[2+d-z]f_6(y,z)\right]
\label{impi:inter}
\eqa
where 
\bqa
\no f_4(y,z)& = &\frac{1}{2}\left[ \log \left\vert \frac{c}{4}\right\vert -\frac{z+y^2}{2z}\log \vert 1+z\vert \right.\\
 & - &\left.\frac{\sqrt{c^2+z^2}}{2z}\log \left\vert \frac{\sqrt{c^2+z^2}-z}{\sqrt{c^2+z^2}+z}\right\vert\right.\\
\no & + & \left. \Theta[y-1]\frac{z+y^2}{2z}\log\left\vert\frac{1+z}{z+y^2}\right\vert\right]
\label{eqn:f4}
\eqa
\bqa
\no f_5(y,z)& =& \frac{1}{2}\left[\log \left\vert\frac{2+z+\sqrt{(2+z)^2+c^2}}{c}\right\vert+\frac{\sqrt{c^2+z^2}}{2z}\right.\\
 & &\hspace{-0.2cm}\left.\log \left\vert\frac{\sqrt{c^2+z^2}(2+z)-z\sqrt{(2+z)^2+c^2}}{\sqrt{c^2+z^2}(2+z)+z\sqrt{(2+z)^2+c^2}}\right\vert\right] 
\label{eqn:f5}
\eqa
\bqa
\no f_6(y,z)&= &\frac{|z-y^2|}{2z}\left[ \frac{\pi}{2}-\tan^{-1}\left(\frac{|z-y^2|(2-z)}{z\sqrt{d^2-(2-z)^2}}\right)\right]\\
& & -\frac{1}{2}\left[\frac{\pi}{2}-\sin^{-1}\left(\frac{2-z}{d}\right)\right]
\label{eqn:f6}
\eqa
\begin{figure*}[t!]
\includegraphics[scale=0.2]{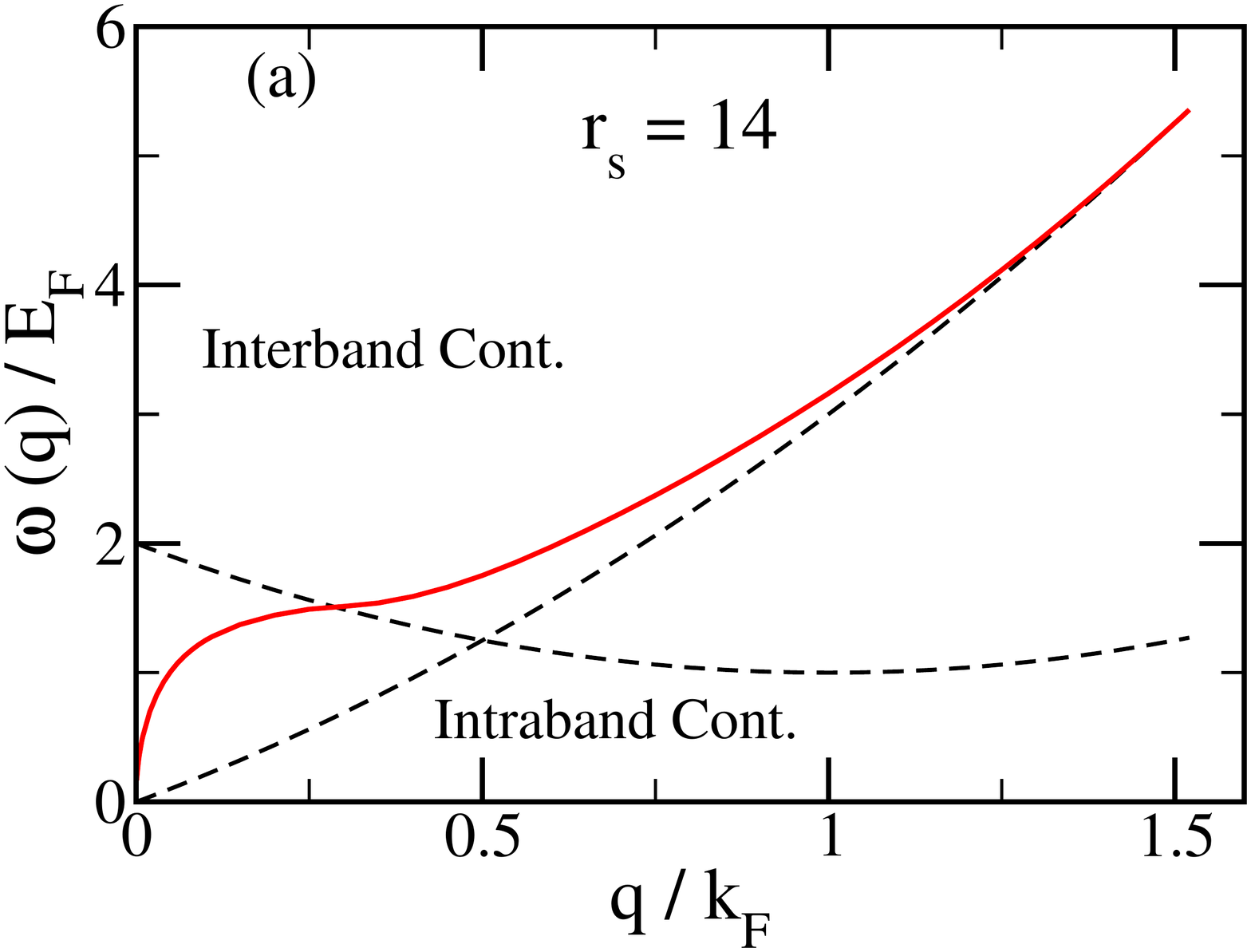}
\includegraphics[scale=0.2]{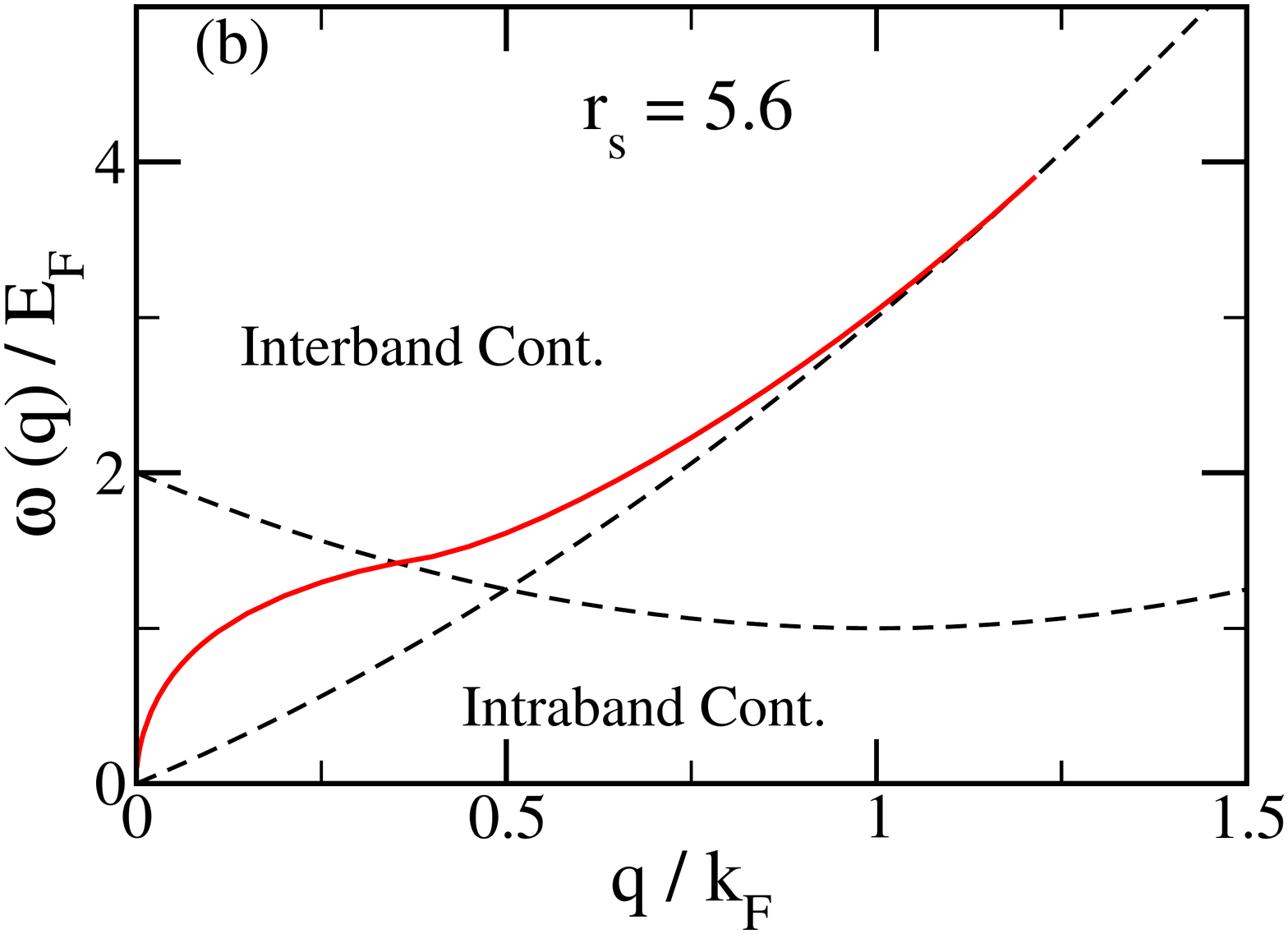}
\includegraphics[scale=0.2]{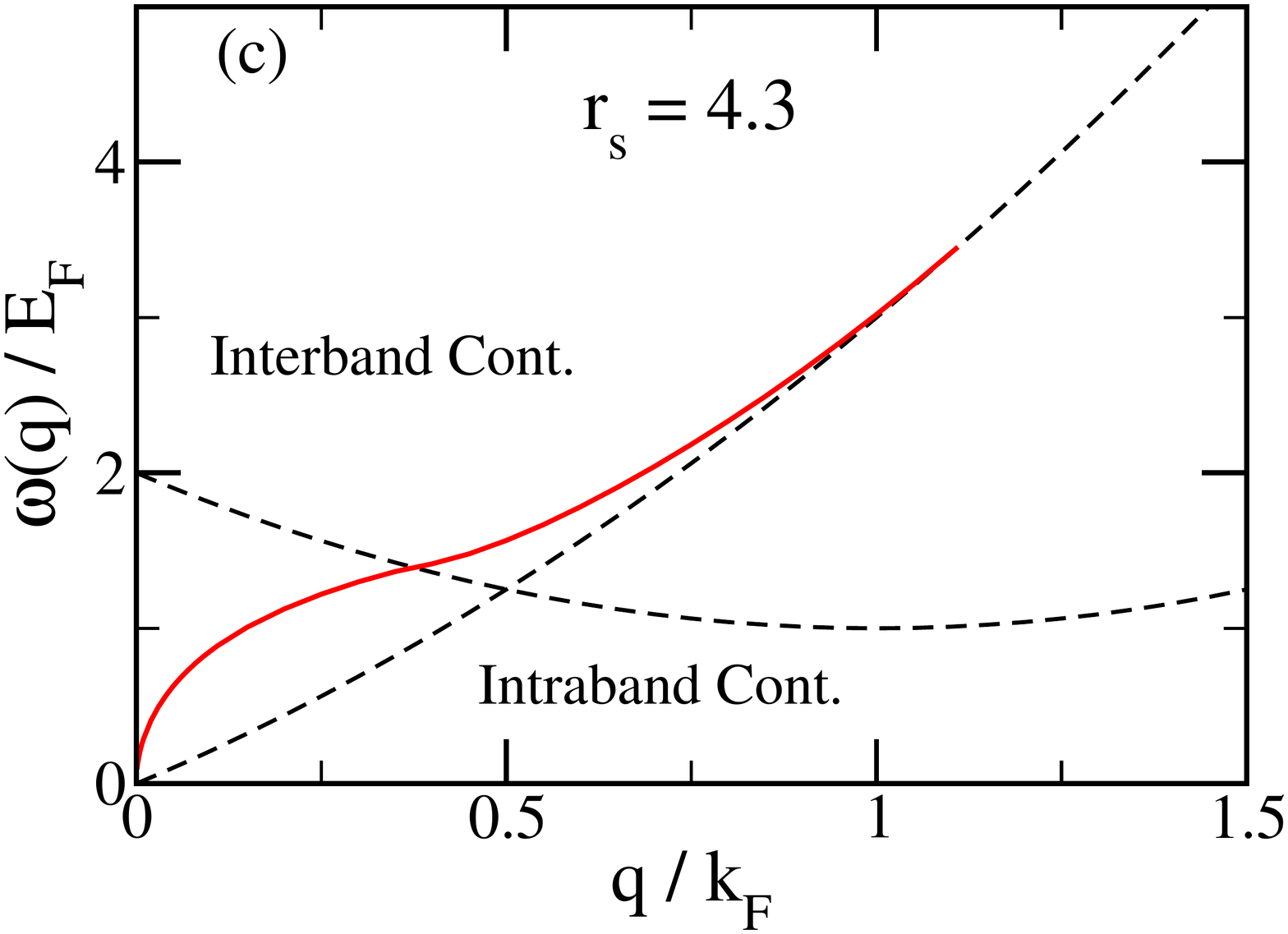}
\caption{Optical plasmon dispersion in bilayer graphene systems for
  (a) $r_s=14$ (b) $r_s=5.6$ and (c) $r_s=4.3$. The thick red line is
  the plasmon dispersion, while the dashed lines show the upper edge
  of the intraband continuum and the lower edge of the interband continuum.
  (color online).}
\label{fig:plasmadisp}
\end{figure*}
Eqn.s (\ref{repi:intra}-\ref{eqn:f6}) are the main analytic results of
this paper. These equations completely define the BLG RPA dynamic
response (and consequently the collective plasmon modes) analytically
within the 2-band low energy quadratic band dispersion
approximation. We first point out the essential features of the
polarizability which differentiate it from the 2DEG and SLG systems.

The intraband polarizability has an imaginary part for $0<z<y^2+2y$
for $y<2$ and for $y^2-2y<z<y^2+2y$ for $y>2$, which defines the band
of particle-hole continuum. For $y<1$, there is a discontinuity in the
derivative at $z=2y-y^2$, which is qualitatively similar to the 2DEG
results. For $2>y>1$, an additional derivative discontinuity appears at
$z=y^2$, which is a result of the chiral nature of the
wavefunctions. For $y>2$, the feature at $z=y^2$ persists, while
$z=2y-y^2$ shifts to the lower edge of the particle-hole
continuum. These features are shown in Fig.~\ref{polintra}(a), (b) and (c)
respectively. These features result in sharp features in the real part
of the polarizability. However, unlike SLG systems, the polarizability
does not diverge at the upper edge of the continuum. The singularity
in this case is softened to a derivative discontinuity. The real part
of the polarizability for different wavevectors is plotted in
Fig.~\ref{polintra} (d), (e) and (f) respectively.
\begin{figure*}[t!!]
\includegraphics[scale=0.2]{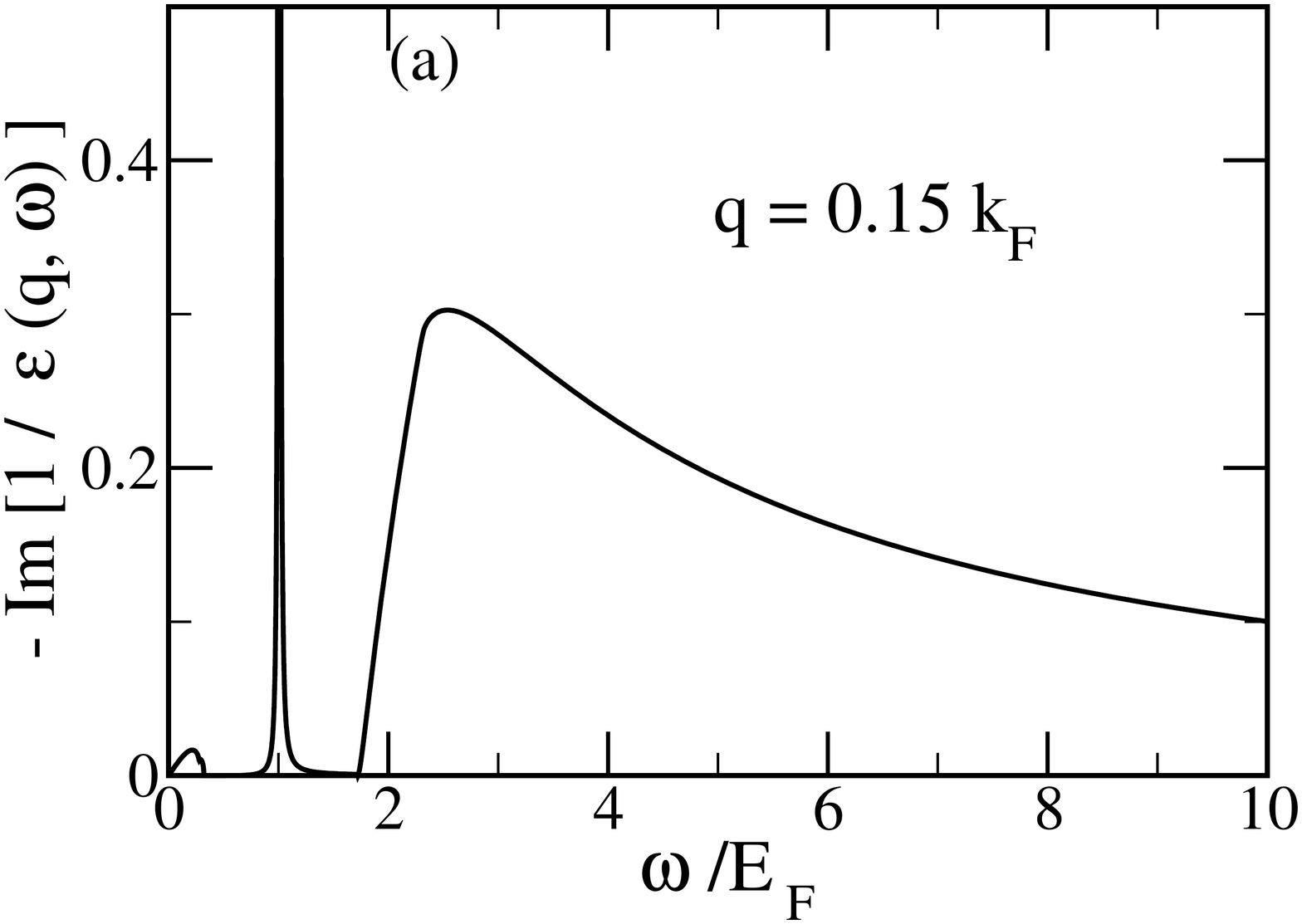}
\includegraphics[scale=0.2]{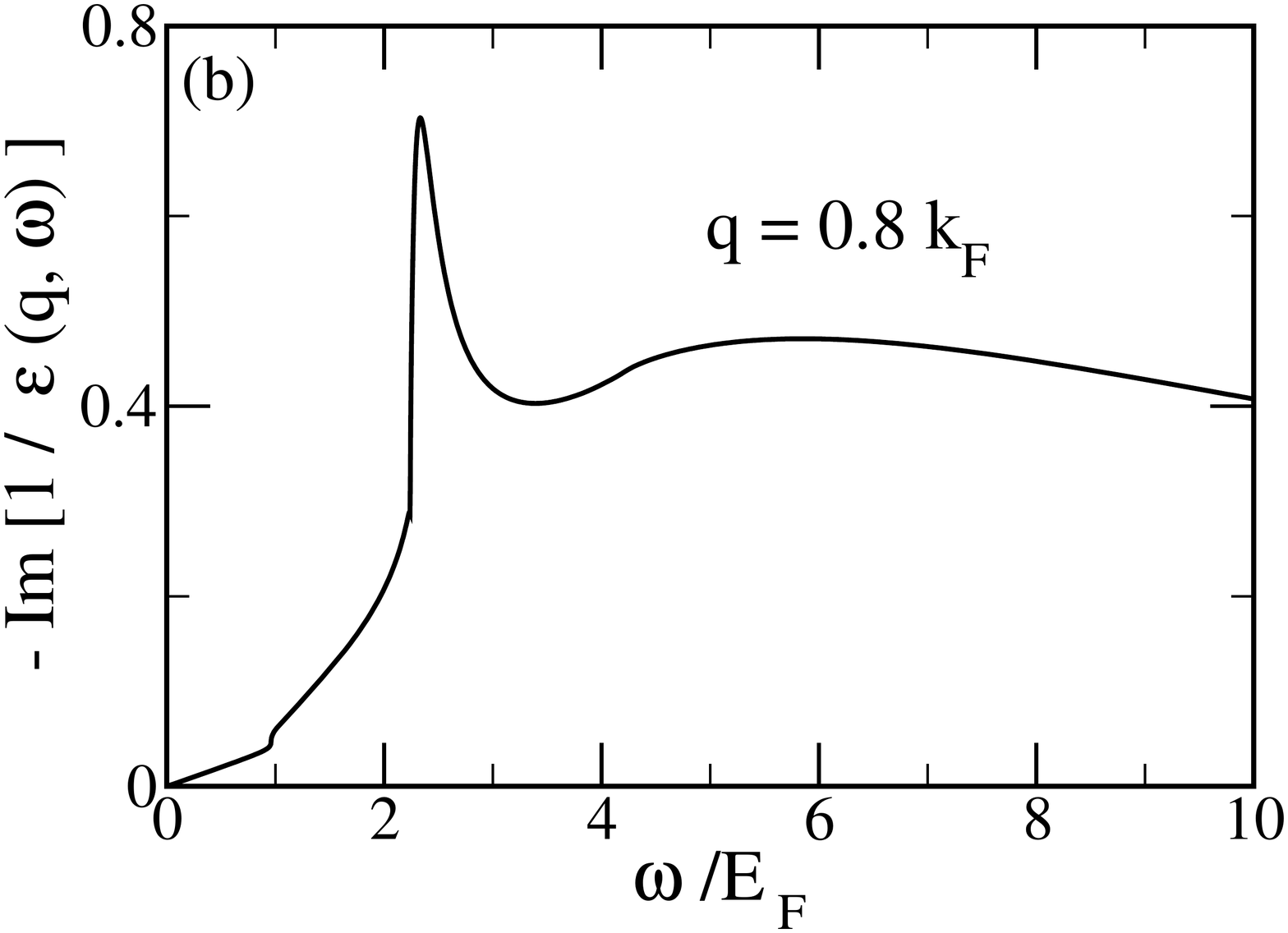}
\includegraphics[scale=0.2]{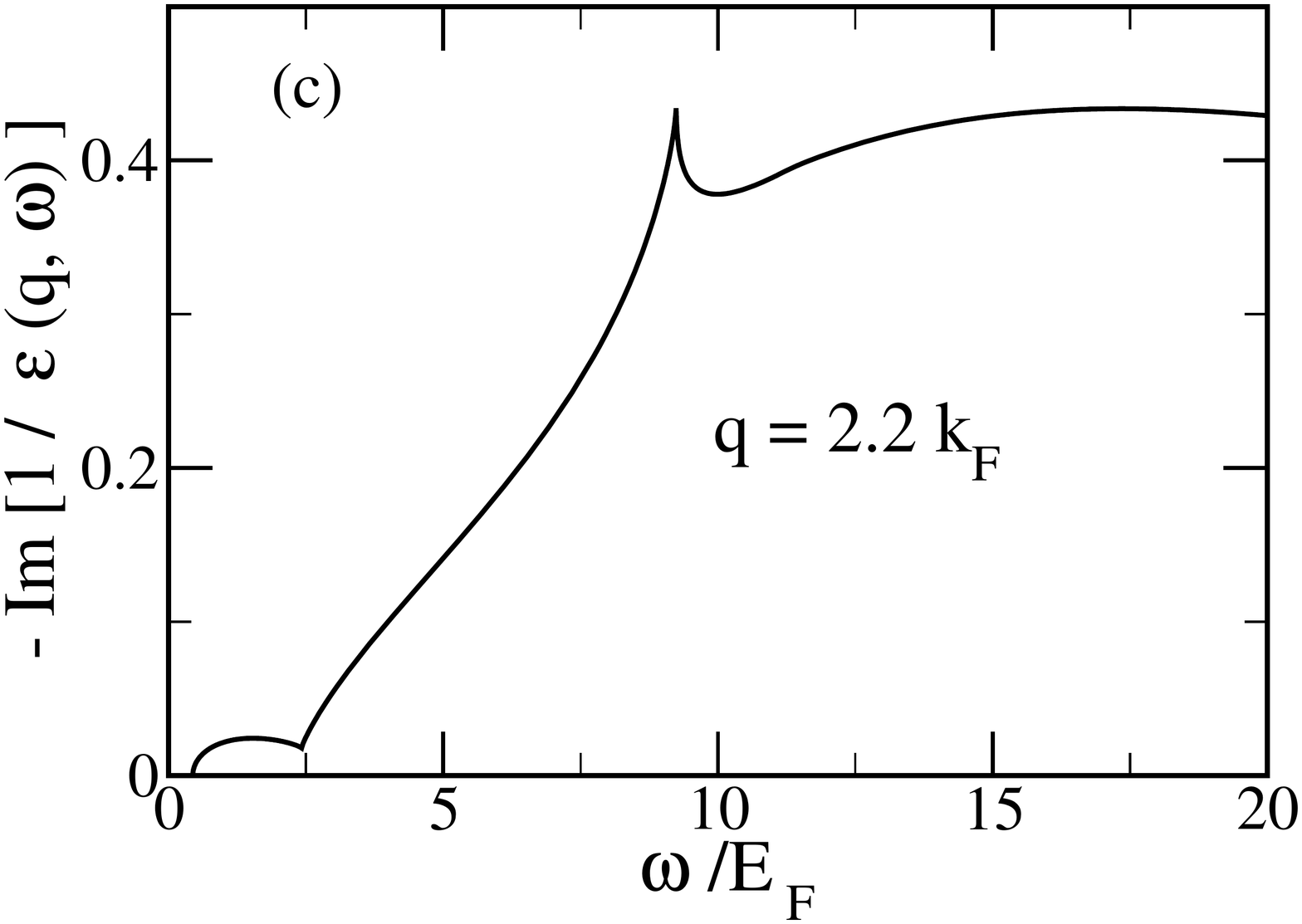}
\caption{ Imaginary part of inverse dielectric function for (a) $q=0.15k_F$ 
(undamped plasmon), (b) $q=0.8k_F$ (damped plasmon) and (c) $q=2.2k_F$ (no plasmon pole). All figures are for $r_s=4.3$.}
\label{fig:optspwt}
\end{figure*}

The imaginary part of the interband polarizability is non-zero for
$z>y^2-2y+2$ for $y<2$ and for $z>y^2/2$ for $y>2$ which defines the
interband continuum. For $z>y^2+2y+2$, it has the simple form $\sim
y^2/z$.  For $y>1$, there is a derivative discontinuity at $z=y^2$. The
real part of the polarizability shows a peak at the lower edge of the
interparticle continuum. The imaginary and real part of the interband
polarizability are plotted in Fig.~\ref{polinter}(a), (b), (c) and
Fig.~\ref{polinter} (d), (e) and (f) respectively.

\section{Plasmons and Optical Spectral weight}

The dielectric function can be used to calculate the dispersion and
damping of plasmons, which are the long-wavelength longitudinal
collective modes of the density fluctuations. The dispersion can be
obtained by looking for poles of the two-particle Green functions or
equivalently for zeroes of the real part of the dynamical dielectric
function. It is useful to express the dielectric
function in terms of the dimensionless quantities $y=q/k_F$ and
$z=\omega/E_F$.
\beq
\ep(\qq,\omega)=1-\frac{r_s}{y}(\Pi^{intra}(y,z)+\Pi^{inter}(y,z)),
\eeq 
where the dimensionless ratio $r_s=e^2gm/(\kappa k_F)$ denotes the
ratio of the interaction energy scale to the kinetic energy scale in
the system. It is important to note that unlike SLG systems, where $r_s$ is 
only a function of the substrate dielectric constant and can take a few 
discrete values, $r_s$ for BLG systems can be varied continuously by changing 
the gate voltages and hence density of the carriers. The coupling constant
$r_s$ in BLG is thus similar to that in 2DEG with carrier density being the 
tuning parameter ($r_s \sim n^{-1/2}$).

There are two collective modes of this system, an optical plasmon mode
which lies above the intraband electron-hole continuum and an acoustic
plasmon mode which lies within the intraband continuum. The acoustic
mode, which has a linear dispersion at long wavelength limit is always
overdamped. Thus it carries very little spectral weight and is not
experimentally relevant. In the following we will focus on the optical
plasmon mode (henceforth referred to as the plasmon).

To get the dispersion of the optical plasmon, we note that the mode
lies between the intraband and interband continuum in the low
wavelength limit. In this limit we can expand the polarizability to
obtain
\beq 
\Pi(y,z) = 2\frac{y^2}{z^2}-\frac{y^2}{4}.
\eeq 
In the long wavelength limit, we find that the plasmon has a leading
order $\sim \sqrt{q}$ dispersion. In this limit, we recover the 2DEG
result in leading order in $q$
\beq 
\omega_q \simeq e\sqrt{\frac{gE_F}{\kappa}}\sqrt{q}.
\eeq 
Expanding to next order, we can obtain the quantum (non-local)
corrections to the plasmon dispersion. The BLG plasmon dispersion is
upto this order is given by
\beq 
\omega_q \simeq e\sqrt{\frac{gE_F q}{\kappa}}\left(1-\frac{q_{TF} q}{8k_F^2}\right).
\eeq 
where $q_{TF}=r_sk_F$ is independent of the density.

We next compare the plasmon dispersion for BLG systems with those of
SLG and a 2DEG. For the sake of comparison, we first note that the
plasmon dispersion formulae for both SLG~\cite{hwang:slgdynamic} and
2DEG~\cite{2DEG} can be cast in a form similar to BLG,
$\omega_q=\omega_0q^{1/2}(1-\alpha q/k_f^2)$. We first focus on the
leading order $\sim \sqrt{q}$ term. For BLG , the coefficient of the
leading order term is given by $\omega_0=(ge^2E_F/\kappa)^{1/2}$,
while for SLG it is given by
$\omega_0=(ge^2v_Fk_F/2\kappa)^{1/2}$. For the 2DEG it is given by
$\omega_0=(ge^2E_F/\kappa)^{1/2}$.  The density dependence ($\sim
n^{1/2}$) of $\omega_0$ in BLG is different from that of
SLG~\cite{hwang:slgdynamic} ($\sim n^{1/4}$), and is a consequence of
a constant as opposed to a linear density of states. The long
wavelength BLG plasmon is thus identical to the ordinary 2D plasmon
and is thus different from the long wavelength SLG plasmon. In
particular, the long wavelength BLG plasma frequency, in contrast to the SLG
plasma frequency \cite{dassarma2009}, is classical and does not have any
$\hbar$ in the leading order dispersion.
\begin{figure}
\includegraphics[scale=0.35,viewport=50 100 500 540,clip]{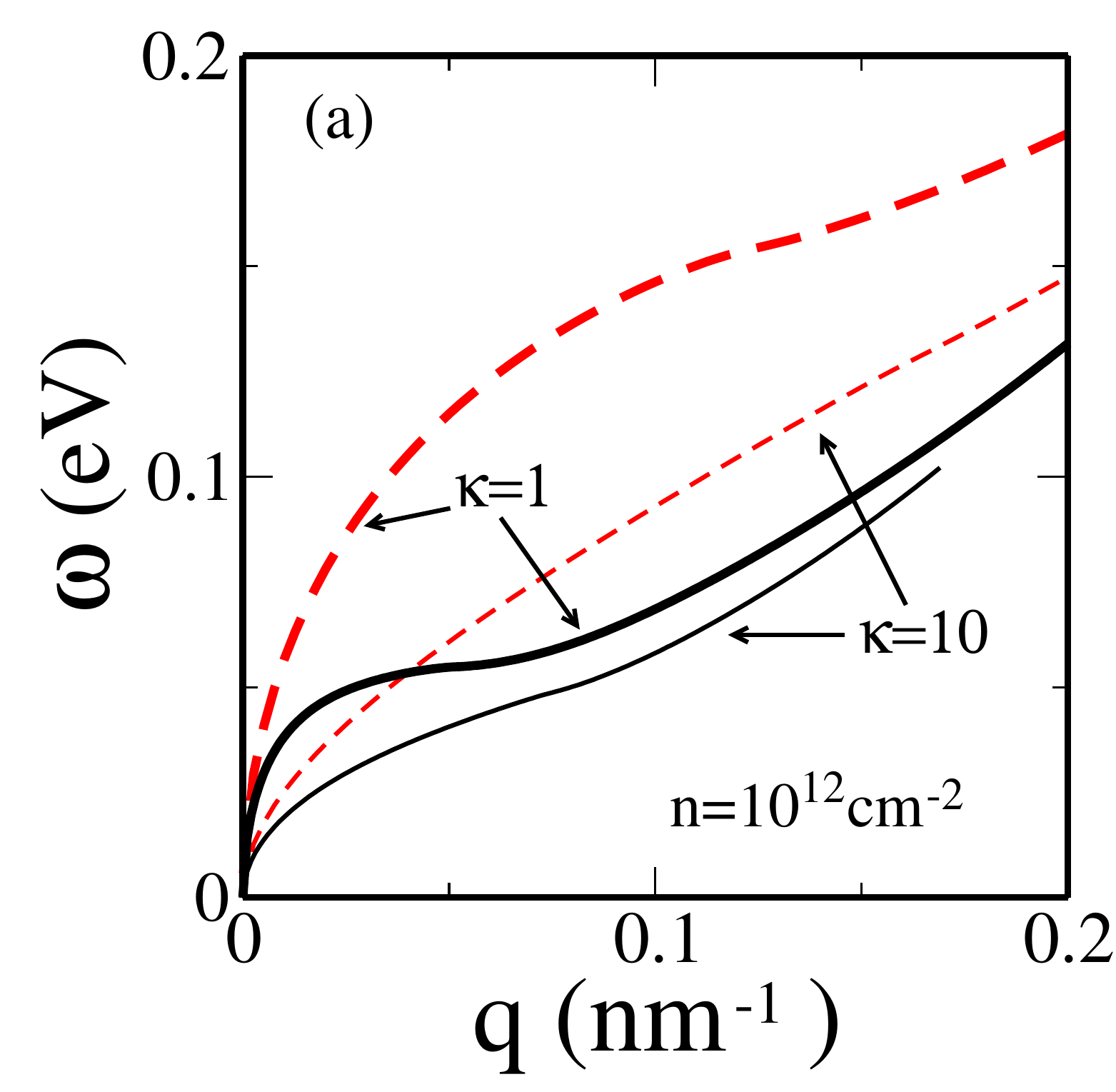}
\includegraphics[scale=0.35,viewport=50 100 500 540,clip]{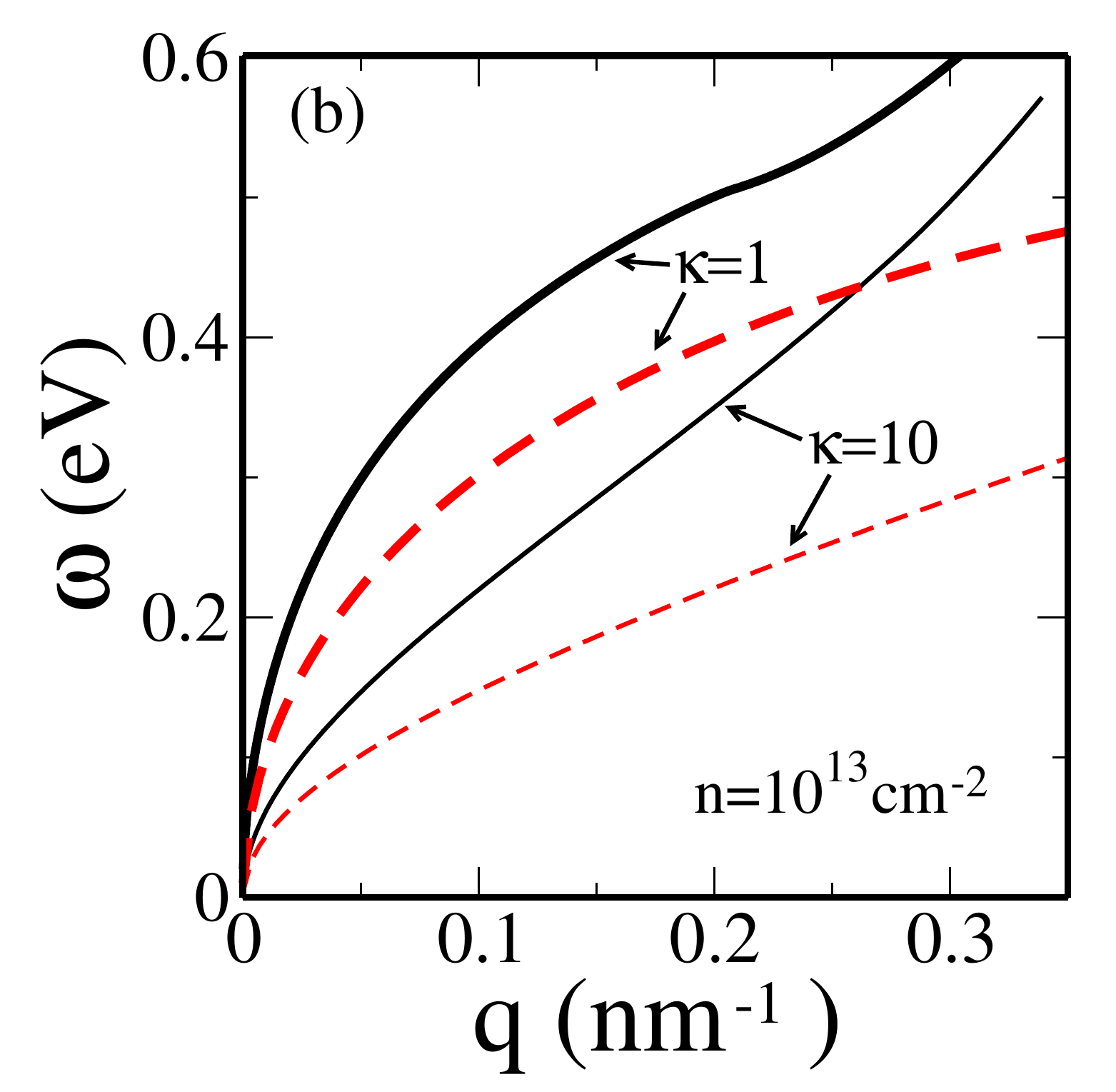}
\caption{Plasmon frequency for BLG (solid black) and SLG (dashed red) systems for two different values of substrate dielectric constant $\kappa=1$ and $\kappa=10$. The two figures are for different densities (a) $n=10^{12}cm^{-1}$ and (b) $n=10^{13}cm^{-1}$ (color online).}
\label{slgcomp} 
\end{figure}

However, the difference with the 2DEG result occurs in the next order
which encapsulates the effect of the quantum corrections. For the BLG
plasmon, $\alpha=q_{TF}/8$ with $q_{TF}=r_sk_F$ being independent of
density. For the 2DEG, $\alpha=-3q_{TF}/4$ with $q_{TF}=gme^2/\kappa$.
Thus, the quantum corrections increase the plasmon frequency from its
leading order result in 2DEG, while in BLG they lead to a reduction of
the plasmon frequency. For SLG, $\alpha=q_{TF}/8$ with
$q_{TF}=ge^2k_F/v_F\kappa\sim n^{1/2}$. The quantum corrections thus decrease 
the SLG plasmon frequency. However the density dependence of this correction 
term is very different from the BLG case.

 The other major difference from the
single layer graphene is that in SLG systems, there is a singularity
of the real part of the polarizability at the upper edge of the
intraband continuum. Hence, the plasmon dispersion never goes into the
continuum and the pole survives upto arbitrary large wave-vectors. In
bilayer systems, there is no discontinuity at the upper edge of the
intraband continuum (it is replaced by a discontinuity in the
derivative with respect to frequency), and so, at a critical
wave-vector, the plasmon pole vanishes once the plasmon dispersion
touches the edge of the continuum.

In order to compare BLG and SLG plasmon dispersions we plot in
Fig.~\ref{slgcomp} the plasmon dispersion (in absolute units of eV) as
a function of wavevector (in absolute units of $nm^{-1}$) for two
fixed density of careers (a) $n=10^{12}cm^{-1}$ and (b)
$n=10^{13}cm^{-1}$ . For each value of density we plot the plasmon
dispersion for two values of $\kappa=1$ and $\kappa=10$. We see that
the BLG plasmon merges with the intraband continuum at a finite
wavevector and gets overdamped while the SLG plasmon is present at all
wavevectors. At the lower density the SLG plasmon frequency is higher
than the BLG plasmon while at the higher density the BLG plasmon has a
higher frequency than SLG plasmon as $\omega_0\propto \sqrt{E_F}$ and
the Fermi energy has different dependence on density in the two cases.

At long wavelengths, the plasmon mode is completely undamped. As the
dispersion goes into the interband continuum at larger wavevectors,
two things happen: (i) the mode gets damped due to usual Landau
damping by the particle-hole pairs, and (ii) the dispersion develops a
knee like structure due to mode repulsion from the continuum. 

In order to look at typical BLG plasmon dispersions, we consider three
types of substrates: (a) $SiC$, which has $\kappa=5.5$, (b) $SiO_2$,
which has $\kappa=4$ and (c) vacuum, corresponding to suspended
bilayer graphene, with $\kappa=1$. Note that the effective $\kappa$ to
be used for calculation of $r_s$ is the average of $\kappa$ for the
substrate and that for the vacuum. For a typical density of $10^{12}
cm^{-2}$, we get $r_s\sim 4.3$ for $SiC$, $r_s=5.6$ for $SiO_2$ and
$r_s=14$ for suspended bilayer graphene. The plasmon dispersion for
these three values of $r_s$ are shown in
Fig.~\ref{fig:plasmadisp}. The critical wavevector for plasmon damping
does not change much varying from $q_1=0.29 k_F$ for $r_s=14$ to
$q_1=0.38 k_F$ for $r_s=4.3$. The maximum wavevector for which the
plasmon pole exists varies from $q_{max}=1.12k_F$ for $r_s=4.3$ to
$q_{max}=1.52k_F$ for $r_s=14$.
\begin{figure}[h!]
\includegraphics[scale=0.2]{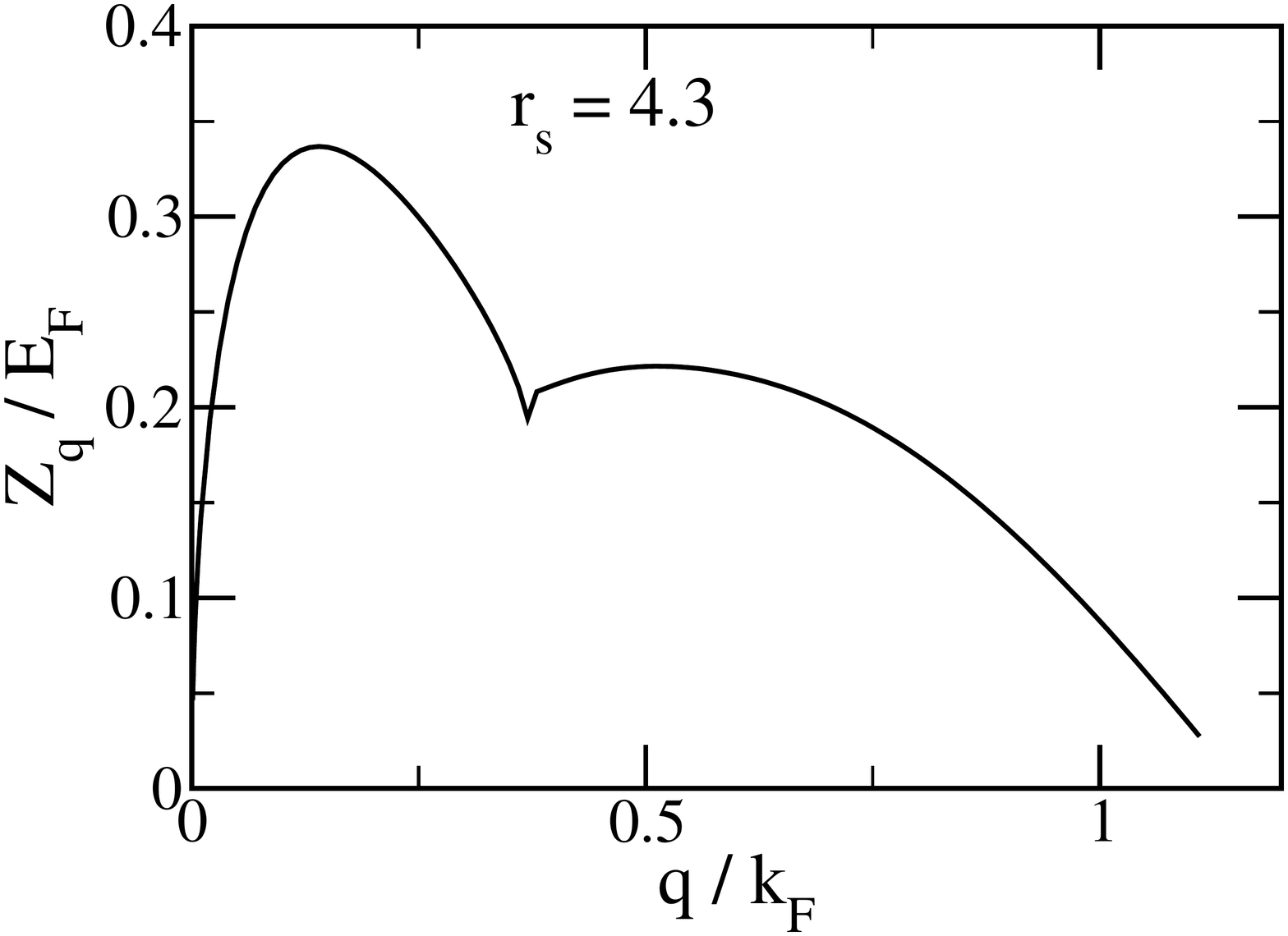}
\includegraphics[scale=0.2]{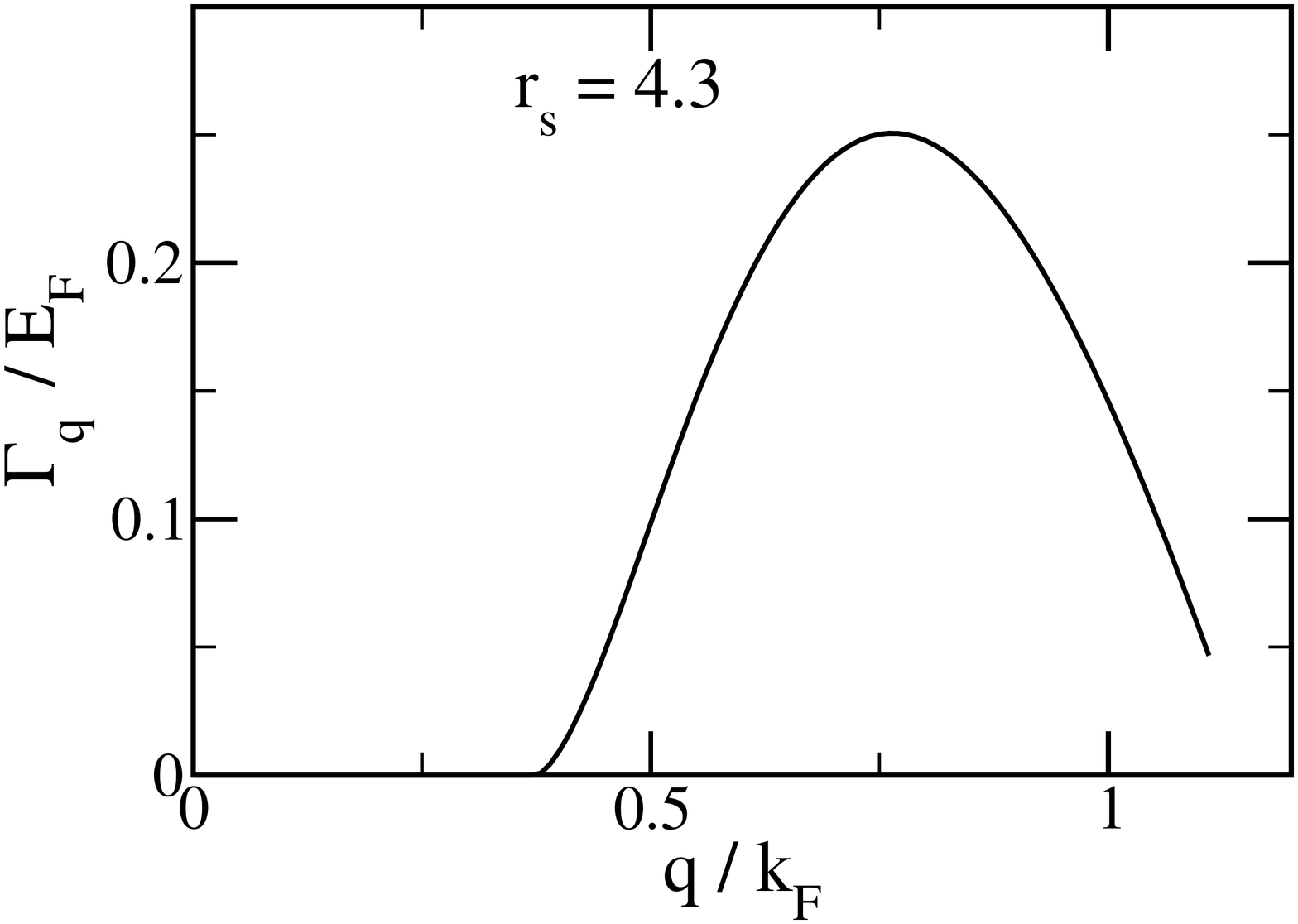}
\caption{(a) The quasiparticle residue $Z$ and (b) The width $\Gamma$ of the 
plasmon pole in the inverse dielectric function for $r_s=4.3$.}
\label{zwidth}
\end{figure} 

An important quantity which is directly measured in spectroscopic
experiments is the optical spectral weight or the loss function. This
quantity is proportional to the imaginary part of the inverse
dielectric function of the system.  In Fig.~\ref{fig:optspwt}, we plot
the spectral weight for three different wavevectors for $r_s=4.3$. The
first wavevector ($q=0.15k_F$) corresponds to the situation where the
plasmon mode lies between the intraband and interband continuum. The
plasmon in this case is sharp, i.e. not Landau-damped. The second
wavevector ($q=0.8k_F$) corresponds to the situation where the plasmon
peak has entered into the interband continuum. In this case we see a
broadened plasmon peak. The third wavevector ($q=2.2 k_F$) corresponds
to the case where the plasmon pole has vanished, and the collective mode is overdamped. It is to be noted
that since the dielectric function continues to have a kink at the
edge of the intraband continuum even when the plasmon has ceased to
exist, a peak like structure is seen in the optical spectral weight in
this limit. This peak vanishes at much larger wavevectors $\sim 4k_F$
for $r_s=4.3$.

Focusing on the plasmons, one can expand the inverse dielectric function near
the plasmon pole to write
\beq
\frac{1}{\ep(\qq,\omega)}=\frac{Z_q}{\omega-\omega_q+ i \Gamma_q}.
\eeq
where the $\omega_q$ is the plasmon frequency, $Z_q=-1/[V(q)\partial
\Pi^{'}(q,\omega)/\partial \omega\vert_{\omega=\omega_q}]$ is the
plasmon pole residue or equivalently, the weight of the plasmon peak,
and the width of the peak is given by
$\Gamma_q=-V(q)\Pi^{''}(q,\omega_q)Z_q$. Here $\Pi^{'}$ and $\Pi^{''}$
denote the real and imaginary part of the polarizability.  In
Fig.~\ref{zwidth} we plot $Z_q$ and $\Gamma_q$ as a function of $q$
for $r_s=4.3$.

The plasmon weight $Z_q\sim \sqrt{q}$ in the long wavelength limit. It
shows a broad peak before the plasmon dispersion enters the interband
continuum. At the lower edge of the interband continuum, $Z_q$ shows a
kink as a function of $q$. It flattens out over a substantial range of
wavevectors before finally going to zero as the plasmon approaches the
intraband continuum and vanishes. The width of the plasmon peak starts
from zero at the interband edge and shows a broad feature before going
to zero again as the plasmon vanishes. The plasmon features predicted
in our theory (Figs. ~\ref{fig:plasmadisp}-~\ref{zwidth}) should be
directly observable in inelastic electron energy loss (IEEL) \cite{slgexpt,ieel} or
inelastic light scattering (ILS) spectroscopies~\cite{lightscatter}
where the loss function can be mapped out as a function of $\qq$ and
$\omega$.

\section{Hyperbolic dispersion and Crossover from BLG to SLG physics}
\begin{figure}
\includegraphics[scale=0.4,viewport= 70 120 500 500,clip]{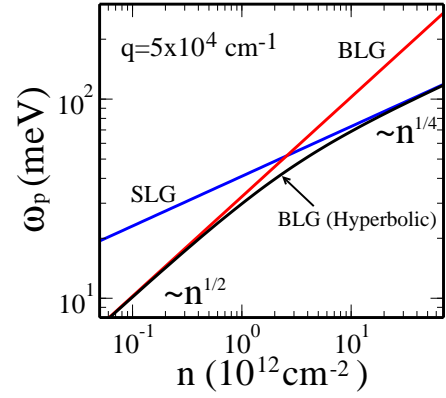}
\caption{The scaling of the leading order plasmon dispersion in the long wavelength limit with density on a log-log scale. The leading order result is obtained by evaluating the plasmon dispersion at a wavevector $q=5\times 10^4 cm^{-1}$. The red line is the BLG result with parabolic dispersion. The blue line is the SLG result. The black line is the BLG plasmon dispersion with hyperbolic dispersion (color online).} 
\label{hyperbolic}
\end{figure}
We have treated bilayer graphene as a system with quadratic band
dispersion and chiral electron and hole bands. While the parabolic
dispersion is relevant at low energies, the actual dispersion of the
bands is hyperbolic. Thus at large wavevectors (relevant at large
densities) the dispersion is effectively linear and the system should
behave like single layer graphene, while at small wavevectors
(relevant at low densities) it should behave like the bilayer graphene
system we have described. In this section, we wish to study this
crossover by focusing on the long-wavelength limit of the plasmon
dispersion. Although the full plasmon dispersion at all wavevectors
can be obtained numerically, working with the leading order long
wavelength dispersion lets us obtain analytic results and hence
provides insight into the crossover. This will also let us estimate
the regime of validity of our parabolic band approximation.

To estimate the long wavelength plasmon, we note that for $qv_F/\omega \ll 1$, the 
bare bubble polarizability is given by
\beq
\displaystyle \Pi(\qq,\omega)=\frac{gk_F}{4\pi}\frac{q^2}{\omega^2}\left. \frac{\partial E_\kk}{\partial_\kk}\right\vert_{k=k_F}
\eeq
where $E_\kk$ is the dispersion of the band containing the Fermi
level. Hence the long wavelength plasmon dispersion is given by
\beq
\omega_q^2=\frac{ge^2k_Fq}{2\kappa}\frac{dE_F}{dk_F}=\frac{ge^2n}{\kappa}\frac{dE_F}{dn}q
\eeq
where $n$ is the density of carriers and we have used $n=gk_F^2/(4\pi)$.
The above results are true for any dispersion including (a) a parabolic
dispersion, (b) a linear dispersion and (c) a hyperbolic dispersion.
Let us now focus on the hyperbolic dispersion of BLG. In this case one can 
write the Fermi energy as a function of density as~\cite{slgreview}
\beq
E_F=\left[\frac{\gamma^2}{2}+\frac{4\pi v_F^2}{g}n-\frac{\gamma^2}{2}\sqrt{1+\frac{16\pi v_F^2n}{g\gamma^2}}\right]^{1/2}
\eeq
where $\gamma$ is the interlayer tunneling and $v_F$ is the Fermi
velocity of the graphene sheets. Note that in the limit of small
$n\rightarrow 0$, $E_F\sim k_F^2/2m$ with $m=\gamma/(2v_F^2)$ and we recover 
the parabolic Fermi energy, while in the large density limit ($n\rightarrow \infty$), $E_F=v_Fk_F$ and we recover the single layer graphene results. Now the long wavelength plasmon frequency for hyperbolic dispersion is then given by
\beq
\omega_q^2=\frac{2\pi ne^2}{\kappa}\frac{v_F^2}{E_F}q\left[1-\frac{1}{\sqrt{1+\frac{16\pi v_F^2n}{g\gamma^2}}}\right]
\eeq
In the low density limit ($n \rightarrow 0$) this reduces to
$\omega_\qq^2=(2\pi ne^2)q/(\kappa m)$ and we recover the leading
order result for the bilayer graphene with parabolic dispersion. In
the large density limit ($n\rightarrow \infty$), the limiting
behaviour is given by $\omega_\qq^2=e^2v_F\sqrt{g\pi n}q/\kappa$ which
matches with the result for single layer
graphene~\cite{hwang:slgdynamic}. In between the answer smoothly
interpolates between these two limits. 

In Fig.~\ref{hyperbolic} we plot the plasmon frequency at a low
wavevector ($q=5\times 10^4 cm^{-1}$) as a function of density on a
log-log scale. The low wavevector ensures that we are looking at the
leading order result. We plot the results for BLG and SLG which show
the usual $n^{1/2}$ and $n^{1/4}$ scaling respectively. The curve for
hyperbolic dispersion of BLG smoothly interpolates between the above
two curves. From the figure we can estimate the value of critical
career density where the hyperbolic result differs significantly from
the parabolic result. By comparing the long wavelenght plasma
frequencies in SLG and BLG we obtain the crossover density as
$n_c=m^2v_F^2/\pi$. This correspond to a density $n\sim 3\times
10^{12}cm^{-1}$ for a bilayer effective mass $m\simeq 0.033 m_e$ (see
Fig.~\ref{hyperbolic}). This then gives us the limit of validity of
the parabolic approximation for BLG systems.

\section{Comparison with Other Bilayer Systems}
\begin{figure}
\begin{center}
\hspace{0.9cm}\includegraphics[width=0.7\columnwidth,viewport=10 100 600 600,clip]{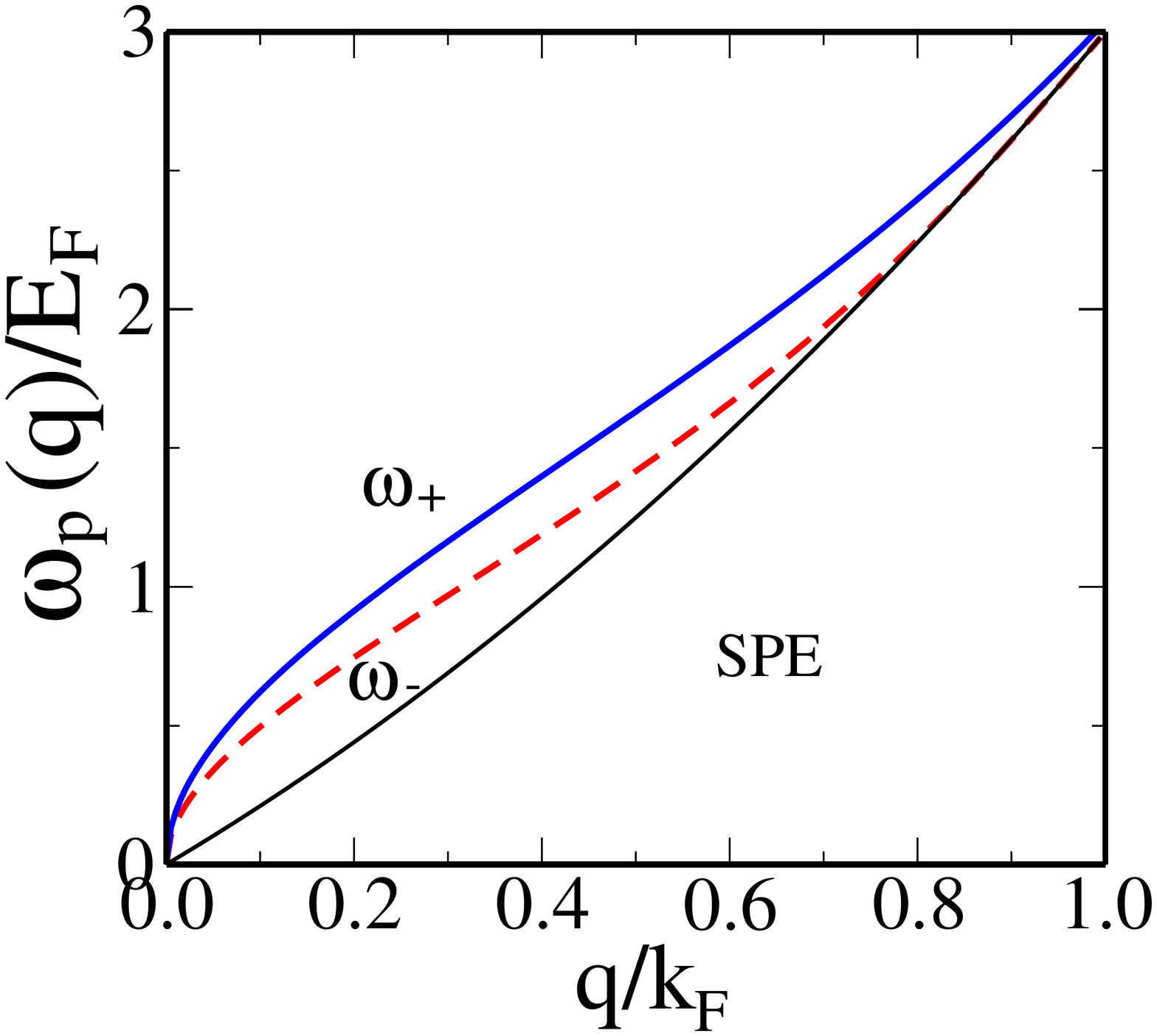}
\includegraphics[width=0.65\columnwidth,viewport=50 30 630 550,clip]{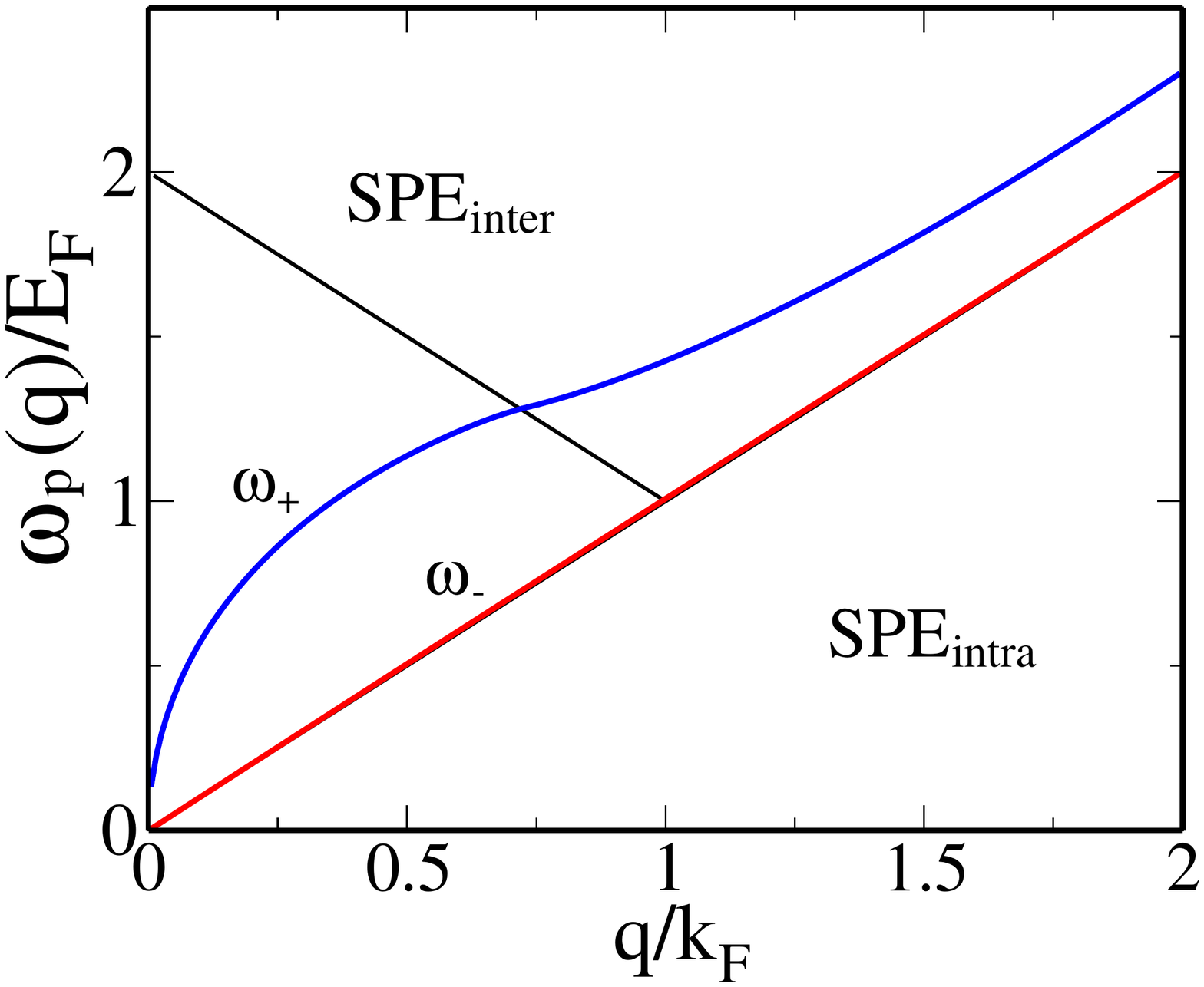}
\includegraphics[width=0.7\columnwidth,viewport=0 0 750 550,clip]{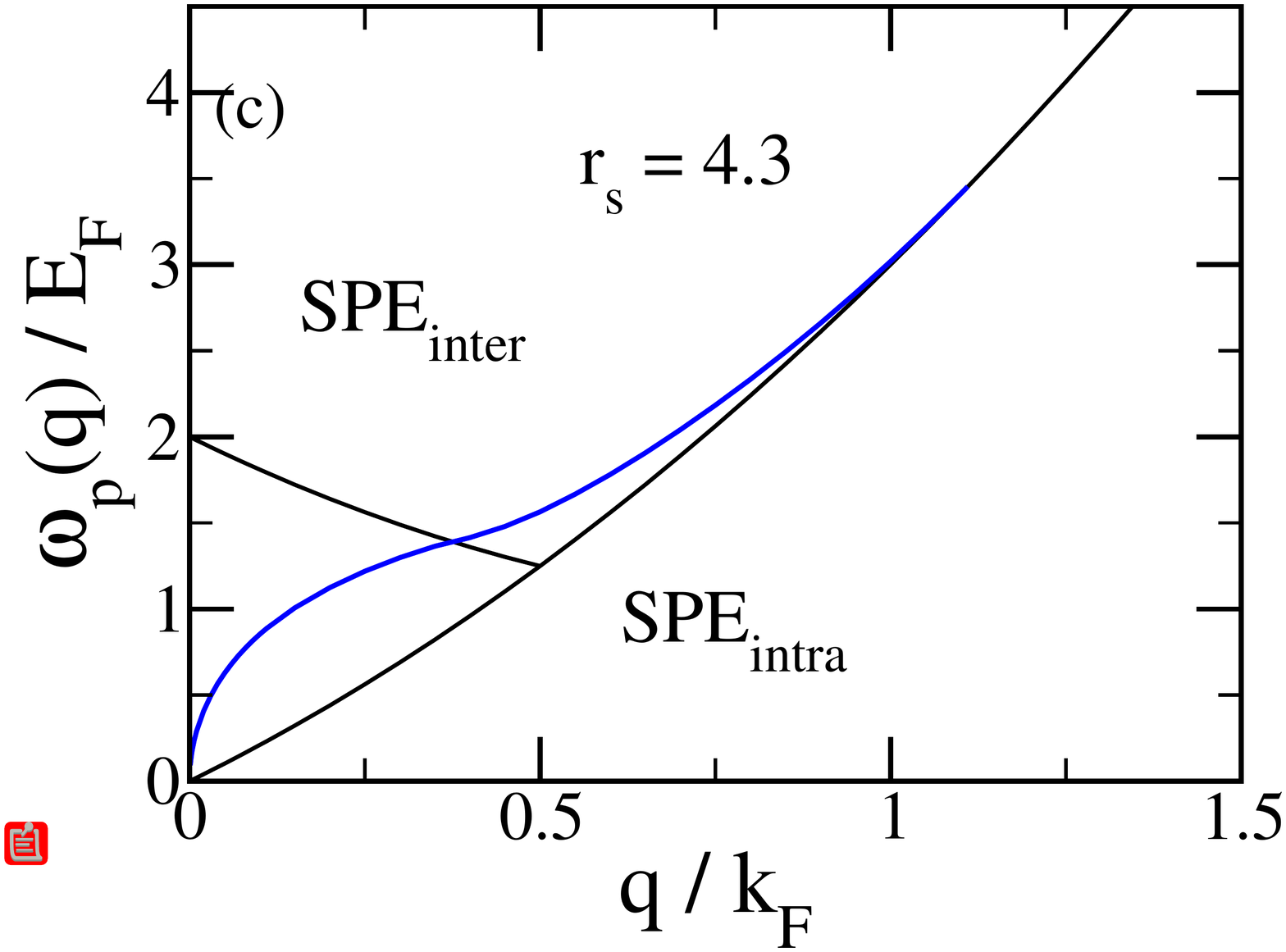}
\caption{Low energy plasmon dispersion for three different bilayer
  systems: (a) an n-GaAs double quantum well with $r_s=1$
  (corresponding to a density of $n=3.2\times 10^{11}cm^{-2}$ and a
  layer separation of $d=9.5 nm$, which corresponds to the effective
  Bohr radius of n-GaAs quantum well) (b) a double layer graphene with
  a density of $n=10^{12} cm^{-2}$ and a layer separation of $d= 3 \AA
  $ (which corresponds to the layer separation of bilayer graphene)
  and (c) a bilayer graphene system with a density of
  $n=10^{12}cm^{-2}$ and $r_s=4.3$. The $sqrt{q}$ mode is in blue and
  the linear mode is in red solid lines. The black lines indicate
  interband and intraband continuum. The undamped linear acoustic mode
  is absent in bilayer graphene ( color online).}
\label{bilayercomp}
\end{center}
\end{figure}

The study of dielectric functions and plasmons for bilayer systems
have a long history starting with the prediction of an undamped
acoustic plasmon mode in double quantum wells~\cite{Madhukar} with
layer separation greater than a critical distance. Similar results
have also been obtained for double layer graphene
systems~\cite{hwang:dlg}. However, the physics of these systems is
quite different from the bilayer graphene system which is of interest
in this paper. These other systems are comprised of well separated 2D
layers coupled only by the Coulomb interaction. Interlayer tunneling is
absent in these systems. On the other hand, the bilayer graphene
system is in the opposite limit of small layer separation with layers coupled strongly through interlayer tunneling,
which is accounted for in the model by changing the low energy
dispersion from linear to quadratic. This leads to all the major
differences between the bilayer graphene and other double layer
systems.

The double layer 2DEG systems in the limit of infinite separation
corresponds to two copies of the underlying 2D systems. Hence it has
two plasmon modes (one in each layer), both of which have $\sim
\sqrt{q}$ dispersion. For long wavelengths, these modes are undamped
as they lie above the particle-hole continuum. The finite separation
and the resultant Coulomb repulsion change the dispersion of one of
the modes to linear in $q$, but with a slope which keeps the plasmon
above the two particle continuum. This is the origin of the undamped
plasmon mode with linear dispersion~\cite{Madhukar}. The slope of the
linear mode decreases with distance, and below a critical layer
separation, the mode gets into the continuum and becomes damped. In
double layer graphene we have the additional feature that the undamped
linear mode overlaps with the optical mode for a range of $q$ values
before splitting off as it enters the interband continuum~\cite{hwang:dlg}.

The crucial missing ingredient in the picture above is the effect of
very strong interlayer tunneling which dominates the physics of
bilayer graphene systems. In the presence of strong interlayer
tunneling, the linear acoustic mode of the interaction coupled bilayer
2D systems is gapped out in BLG. In fact, the two band representation
of BLG systems is an effective low energy description. A more detailed
description~\cite{macdonald} would include $4$ bands, two of which are
gapped out (with gaps $\sim eV$). In this case one would recover the
gapped acoustic plasmon mode sitting at an energy above the highest
band continuum. It is to be noted that the energy of the gapped mode
is $\sim eV$ and therefore does not affect the low energy properties
of the system. Our low energy description of BLG system then recovers
the low energy optical plasmon mode with $\sim \sqrt{q}$ dispersion.

In Fig.~\ref{bilayercomp}, we show in the same figure, for the sake of
comparison, the low energy plasmon dispersion for typical parameter
values for the three bilayer systems. The double 2DEG dispersion is
calculated for large layer separations and clearly shows the undamped
optical and acoustic plasmon modes. The double layer graphene
calculation is done for a relatively shorter layer separation. In this
case the acoustic plasmon mode is quite close to the particle-hole
continuum. Finally, the linear acoustic mode is not seen in bilayer
graphene systems since the BLG, being strongly tunnel-coupled is effectively a 
1-component system similar to a single 2DEG or SLG, but with its own distinct plasmon dispersion.

\section{Conclusions}

In this paper, we have analytically studied the dynamic screening
properties of doped bilayer graphene systems within RPA. We have
obtained analytic forms for the full wavevector and frequency
dependent dielectric function for the system at arbitrary
densities. From this we have obtained the plasmon dispersion and
calculated the plasmon spectral weight, focusing on the plasmon
residue and broadening. We find that while the leading order plasmon
dispersion matches with the classical 2D electron gas result, the
non-local dispersion correction in the next order suppresses the
plasmon frequency, contrary to the 2D electron gas. However, the
density dependence of the Thomas Fermi wavevector is different from a
single layer graphene, which is a consequence of the constant density
of states as opposed to linear density of states in single layer
graphene. This leads to a $n^{1/2}$ density dependence of the BLG
plasma frequency in contrast to the $n^{1/4}$ dependence for SLG.

We have also compared the results for bilayer graphene systems with
those in double quantum wells and double layer graphene systems. We
find that while the Coulomb coupling between the layers results in an
undamped linearly dispersive acoustic plasmon mode in double quantum
wells and double layer graphene systems, the interlayer tunneling,
which dominates the physics in the bilayer graphene systems, gap out
the acoustic mode and we are left with a single undamped optical
plasmon mode (dispersing as $\sqrt{q}$) in the low energy limit.


In conclusion, we discuss the approximations and the limitations of
our theory, and in the process point out possible future research
directions if experimental data on BLG plasmon spectra, when they
become available, necessitate further extension of our theory. We have
used a $T=0$ RPA many-body theory for obtaining BLG dynamic dielectric
response and calculating BLG plasmon spectra (plasmon dispersion,
damping, and spectral weight). We have also assumed a clean system,
neglecting any disorder effect on the dielectric response and the
plasmons. In addition, we have made a 2-band parabolic single-particle
band dispersion approximation. All of these approximations can, in
principle, be improved albeit at the considerable cost of losing the
analytic simplicity of our current theory. 

Given that the BLG Fermi temperature for a given carrier density $n$
is given by $T_F \simeq 400 \tilde{n} K$, where $\tilde{n}= n/
10^{12}cm^{-2}$, our $T=0$ theory should be quite good even at room
temperature for $n > 10^{12}cm^{-2}$ and down to $n\simeq
10^{11}cm^{-2}$ (or perhaps even lower) as long as $T \geq 4 K$. The
disorder effect is fairly easily incorporated in our theory (at least
in the leading order) by interpreting the frequency $\omega$ to be
modified by the substitution $\omega^2 \rightarrow \omega(\omega + i
\Gamma_i)$, where $\Gamma_i\equiv \hbar/(2\tau_i)$ is obtained from
the system mobility $\mu$ ($\equiv \sigma/(ne)$, where $\sigma$ is the
conductivity) by writing $\mu=e\tau_i/m$. The main effect of disorder
is thus to introduce a plasmon damping of $\Gamma_i$ even outside the
electron-hole Landau continuum regime. In the presence of Landau
damping of plasmons, the disorder-induced $\Gamma_i$ adds to the
plasmon broadening arising from Landau damping. In absence of Landau
damping and at low temperatures, the disorder broadening $\Gamma_i$ is
the main mechanism contributing to plasmon damping. 

In contrast to thermal and disorder effects, our other approximations
(i.e. RPA and parabolic dispersion) are not easy to improve upon. The
parabolic band structure approximation of our theory would fail at
high energy where the BLG single-particle dispersion becomes linear,
similar to SLG. This can be incorporated numerically in the RPA
dielectric function, thus sacrificing the analytic transparency of our
results. We refrain from doing so because the SLG plasmon dispersion
and dielectric response have already been calculated in the literature
~\cite{hwang:slgdynamic}. When the BLG Fermi level is high in the
conduction or valence band, the BLG dynamical response and plasmon
mode dispersion will smoothly crossover to the SLG result. This is
expected to happen for a carrier density $\geq 5 \times 10^{12}
cm^{-2}$, where $E_F$ is large enough for the BLG band dispersion to
be better approximated by a linear than a parabolic dispersion.

Finally, RPA ignores many-body corrections such as vertex and
self-energy effects in the polarizability, and includes only the
important long-range divergence of the Coulomb interaction through the
infinite series of bubble (or ring) polarizability diagrams. While RPA
becomes exact for small $q/k_F$ (for all $r_s$) and for all $q/k_F$
(for $r_s \ll 1$), higher order self-energy and vertex corrections
should become important for the polarizability function
$\Pi(\qq,\omega)$ for large $q/k_F$ and large $r_s$. Unfortunately, no
systematic improvement of RPA dynamical dielectric response is
available for arbitrary $r_s$ values since such a theory must somehow
include consistent and conserving infinite series of many body
diagrams. In the literature, unsatisfactory approaches based on
``local field'' corrections (i.e. corrections associated with
correlation effects at large wavevectors transcending the zero
wavevector Coulomb divergence) are often used to go beyond RPA, but
theories involving local field corrections are not consistent from a
diagrammatic many-body viewpoint since many-body diagrams from
different perturbative orders are mixed in an uncontrollable
manner. We can easily incorporate the most popular form of the local
field approximation, the so-called Hubbard approximation \cite{hu}, by rewriting
eqn.~\ref{eqn:dielectric} as 
\beq 
\ep(\qq,\omega)= 1 - \frac{2\pi e^2}{\kappa q}\frac{\Pi(\qq,\omega)}{\left\{1+\frac{2\pi e^2}{\kappa
      q}G(q)\Pi(\qq,\omega)\right\}}. 
\eeq 
where the term in the denominator in curly bracket is the local field
correction due to many-body effects neglected in RPA. The explicit
form for the actual local field term $G(q)$ is somewhat arbitrary
except that $G(q)=0$ for $q=0$, so that the RPA is recovered in the
$q\rightarrow 0$ limit. In 2D,
$G(q)=(1/2)(q/\sqrt{q^2+k_F^2+q_{TF}^2})$ is often used although the
rigorous validity of such a local field term in improving RPA is
unknown. We do not pursue this line of reasoning because such local
field corrections are uncontrolled and static so that it is completely
unclear whether it is an improvement on the dynamical RPA. If future
BLG plasmon experiments show systematic deviations from our RPA
predictions with increasing $r_s$ (i.e. decreasing density, where our
parabolic band approximation becomes better), then going beyond RPA
may become necessary.

Acknowledgement: This work is supported by US-ONR MURI.

\end{document}